\documentclass[pre,aps,twocolumn,10pt,superscriptaddress]{revtex4-2}

\usepackage{amsfonts}
\usepackage{times}
\usepackage{bbm}
\usepackage{amsmath, amsthm}
\usepackage{graphicx}
\usepackage{amssymb, algorithm, algpseudocode}
\usepackage[normalem]{ulem}

\algnewcommand\algorithmicforeach{\textbf{for each}}
\algdef{S}[FOR]{ForEach}[1]{\algorithmicforeach\ #1\ \algorithmicdo}
\usepackage[title]{appendix}
\usepackage{xcolor,comment}

\definecolor{darkgreen}{RGB}{0,180,0} 

\begin{document}

\title{The vortex comb: eliminating vortices from Bose-Einstein condensates using optical lattices}

\author{Shrohan Mohapatra}
\affiliation{Department of Mathematics and Statistics, and 
Department of Physics, 
University of Massachusetts, Amherst, Massachusetts 01003-4515, USA}

\author{Andrew J. Schaffer}
\affiliation{Wyant College of Optical Sciences, 
University of Arizona, Tucson, AZ 85721, USA}
\affiliation{Quantinuum, Golden Valley, MN, USA}

\author{P. G. Kevrekidis}
\affiliation{Department of Mathematics and Statistics, University
of Massachusetts, Amherst, Massachusetts 01003-4515, USA}

\author{R. Carretero-Gonz\'alez}
\affiliation{Nonlinear Dynamical Systems Group, 
Computational Sciences Research Center, and 
Department of Mathematics and Statistics, 
San Diego State University, San Diego, California}

\author{B. P. Anderson}
\affiliation{Wyant College of Optical Sciences, 
University of Arizona, Tucson, AZ 85721, USA}
\date{\today}

\begin{abstract}
In the present work we introduce and explore a technique for the efficient removal of vortices from an atomic Bose-Einstein condensate, through the application and subsequent removal of a one-dimensional optical lattice. We showcase a prototypical experimental realization of the technique that motivates a detailed theoretical study of vortex removal mechanisms. Through simulations of the condensate dynamics during application of the optical lattice, we also discover a vortex removal mechanism that arises in narrow, optical-lattice-induced atomic density channels for which the channel width is on the order of the nominal vortex core size and healing length. This mechanism involves the density profile typically associated with a vortex core spatially separating from the phase singularity associated with the vortex. By analyzing numerical experiments covering a wide range of variations of the optical lattice amplitude and fringe periodicity, we identify the existence of an optimal set of parameters that enables the efficient removal of all vortices from the condensate. This analysis paves the way for further studies aimed at understanding vortex dynamics in narrow channels, and adds to an experimental toolkit for working with vortices and controlling the dynamical states of condensates.  
\end{abstract}

\maketitle

\section{Introduction}
\label{sec:introduction}

Quantized vorticity is a hallmark of superfluidity and superconductivity~\cite{Don1991.Vortices.book,tilley1990superfluidity,pethick,Pitaevskii2003}. In dilute-gas Bose-Einstein condensates (BECs), quantized vortices are readily observable and serve as sensitive indicators of superfluid dynamics~\cite{fetter_svidzinsky_2001,MPLB05,Fetter2}. Moreover, vortices in a BEC can be created, manipulated, and imaged using a variety of experimental techniques~\cite{And2010.JLTP161.574}, enabling precise control and detailed studies of quantum fluid behavior; see for example Refs.~\cite{samson,bettina}. In a wide range of experimental settings such as these, vortices and vortex-vortex interactions~\cite{newton} play quintessential roles in BEC fluid dynamics, and an understanding of their dynamics is essential in studies of complex quantum fluid phenomena such as superfluid turbulence and its associated cascades~\cite{white2014vortices,TSATSOS20161}.

Vortices may also form spontaneously during BEC creation~\cite{Sch2007.PRL98.110402,Wei2008.Nat455.948} or as a consequence of the condensate’s motion relative to an external potential~\cite{Nee2010.PRL104.160401}. In such cases, vortices can act as unwanted defects that hinder the precise control and characterization of the dynamical state of the BEC. While techniques have been explored to remove quantized vortices from superconductors~\cite{veshchunov2016optical}, little attention has been paid to the controlled removal of vortices from a BEC, with the goal of preparing the system in its motional ground state prior to subsequent experimental protocols that involve manipulation and coherent control of the state of the condensate.

Here we introduce and experimentally demonstrate a method for the controlled removal of vortices from a highly oblate BEC by briefly applying a one-dimensional optical lattice (OL) potential, and we present detailed numerical studies elucidating the accompanying vortex dynamics and physical mechanisms responsible for the vortex dynamics that lead to their removal. In addition to established vortex dissipation mechanisms, such as thermal damping~\cite{Fedichev1999,Svidzinsky2000,fetter_svidzinsky_2001,PhysRevA.69.053623,Jackson2009,Freilich2010,Rooney2010,Yan2014,yongilshin} and vortex-antivortex annihilation~\cite{frisch1992transition,Nee2010.PRL104.160401,Kwon2021}, we observe in simulations two additional effects: rapid channeling of vortices to the edge of the condensate by the OL, and the dissociation of vortex phase singularities from the density profiles typically associated with a vortex core. To the best of our knowledge, the latter mechanism, apparently occurring when the OL periodicity is on the order of the spatial width of the vortex, has not been reported previously.

The general concept underlying our vortex elimination method is to use an OL to manipulate the density of the BEC, rendering the BEC density distribution as an array of narrow (yet still mutually coherent) channels of atoms.  The channels, in turn, influence vortex dynamics by causing the vortices to move along the channels to the outer edges of the BEC where they can then become eliminated from the fluid.  To broadly and conceptually describe this process, we consider a highly oblate, nearly two-dimensional (2D) BEC that is trapped in a three-dimensional (3D) harmonic potential, for which the $\hat{z}$ axis corresponds to the direction strong confinement, and $\hat{x}$ and $\hat{y}$ define the weaker trapping axes in the 2D plane.  Vortices in such a system will have positive or negative circulation that is quantized about $\hat{z}$, denoted here as vortices (V) and anti-vortices (AV), respectively.  The motion of every vortex in the quantum fluid is governed both by the distribution of all other vortices and by the condensate’s  density gradients.   As a vortex approaches the edge of the BEC, its velocity component parallel to the BEC boundary increases in magnitude and its trajectory becomes increasingly parallel to the BEC boundary. The vortex may then dissipate at the BEC boundary, a process that is enhanced at higher temperatures of the atomic system.  Similarly, a blue-detuned laser beam propagating along $\hat{z}$, and elongated along $\hat{y}$ and tightly focused along $\hat{x}$ can be used to slice the BEC into two regions and define an interior BEC edge.  A vortex that approaches this interior edge is expected to move along the inner BEC edge induced by the laser beam, and to thereby be induced to move to the outer edge of the BEC whereupon it can be eliminated from the BEC by dissipation.   Surprisingly, there appear to be relatively few studies concerning the motion of vortices in more narrow superfluid regions; see, e.g., Ref.~\cite{Galantucci_Sciacca_Parker_Baggaley_Barenghi_2021}, which tackles vortex leapfrogging in such settings.  

Our envisioned vortex removal technique utilizes a one-dimensional blue-detuned OL to segment the BEC into a series of narrow parallel channels; the segmentation is significant relative to the density distribution of atoms in the BEC, but is not so strong as to eliminate the phase coherence between segments of the BEC.  To the best of our knowledge, such segmentation has not been used in studies of vortex dynamics in parallel, narrow elongated channels.  Accordingly in our setting, any vortices present prior to the application of the OL potential, such as vortices that might be created spontaneously during BEC formation or generated by a stirring process, should be channeled to the edges of the BEC once the lattice potential is applied.  The OL can then be removed, returning the BEC to the harmonic potential in a vortex-free state under appropriate conditions of OL potential depth and periodicity.  We refer to this entire process of vortex removal by application of the OL as a ``vortex comb,'' in analogy with the use of a comb to remove tangles from hair. Our aim, upon showcasing a proof-of-principle experimental demonstration of the concept, is to establish a theoretical and computational backdrop for the vortex comb procedure. Detailed numerical simulations performed under broad ranges of choices for the OL amplitude and periodicity allow us to assess the efficacy of the combing process and to analyze the vortex removal mechanisms that are present in the process.  

The remainder of this paper is organized as follows. We first describe the experimental procedure and present results demonstrating the effectiveness of the vortex comb in Sec.~\ref{sec:experiments}.  We then turn to a theoretical framework presented in  Sec.~\ref{sec:theory}, based on which we conduct numerical simulations described in Sec.~\ref{Sec:numerics}.  The associated findings provide a conceptual understanding of the physical processes observed to be at work in the combing process.  Our simulations reveal that the simple picture of vortex channeling presented above is incomplete; under optimized conditions, we observe more complex dynamics, including the density-phase separation process mentioned above. For the parameter ranges studied, we identify optimal conditions for the combing process, and finally, upon summarizing our results, we provide conclusions and possibilities for future research in Sec.~\ref{sec:conclu}.

\section{Experimental Demonstrations}
\label{sec:experiments}

Our experiments begin with the creation of  a $^{87}$Rb BECs of up to $1.3\times 10^6$ atoms held in a hybrid optical-magnetic trap; see Ref.~\cite{schaffer2021methods} for apparatus and experimental details.  The BECs are created with atoms in the $5\,^{2}S_{1/2}$, $F=1$, $m_F=-1$ hyperfine state.  The trap is highly oblate and approximately harmonic, with measured trap frequencies of ($\omega_x$, $\omega_y$, $\omega_z$) = 2$\pi \times (2.8, 8.5, 38.5)$ Hz where  $\hat{z}$ corresponds to the vertical direction.  Typical chemical potentials for the BECs are approximately $\mu_0 \approx 1.4\times 10^{-31}$ J $\approx 5.6\, \hbar\omega_z$ as calculated from a measured Thomas-Fermi radius of 80 $\mu$m along the $\hat{x}$ direction. From these parameters, the healing length $\xi_0$ at maximum density is calculated to be $\xi_0 \approx 0.52\,\mu$m. Although the BEC is a 3D quantum fluid, it is nevertheless highly oblate, and vortex dynamics are expected to be nearly two-dimensional~\cite{PhysRevA.84.023637}. Accordingly, in this limit, each singly quantized vortex has positive or negative circulation that is quantized about the $\hat{z}$ direction.  As such, in order to determine the presence of vortices, we image the BECs along the $\hat{z}$ axis after approximately 12~ms of expansion following the release of the BEC from the optical component of the hybrid trap; the magnetic component remains present to support the atoms from falling due to gravity.  In the expansion procedure, the BEC expands rapidly along the $\hat{z}$ direction with an accompanying decrease in atomic density, enabling any vortex cores present to expand in size and become
visible in the image of the BEC. 

Vortices are stirred into the BEC by adiabatically ramping on a 660-nm laser beam that pierces the BEC.  The beam is focused to a beam radius of 10 $\mu$m at the BEC.  The maximum optical potential provided by the beam is equivalent to approximately the BEC chemical potential $\mu_0$.  By applying time-dependent bias magnetic fields, the BEC is moved in four small circles in the $(\hat{x},\hat{y})$ plane over 800~ms, such that the laser beam acts as a stirring rod to induce the formation of vortices similar to earlier work of a subset of the authors in connection to quantum turbulence in Ref.~\cite{kodylaw}.  The laser beam is then adiabatically ramped off, leaving behind approximately 18 vortices on average in the BEC.  Our experiments with the vortex comb, described next, begin under these initial conditions. 

Without the comb, after 2~s of additional hold time, the mean number of vortices observed was 7.8~(1.5). With 4~s of additional hold time after stirring, the observed mean vortex number was 6.8~(0.48), and for 8~s of hold time the observed mean vortex number was 5.4~(1.2).  The uncertainties reported in parentheses are the standard deviations of counted vortices obtained through repeated observations after identical stirring and hold processes. The fast initial decay of vortex number is presumably due to a combination of thermal damping with vortices being lost at the edge of the BEC, and to V-AV annihilation.  As we will show, additional vortex dynamics processes play key and distinct roles in the combing protocol relevant when the OL is applied.   

The vortex comb is created by intersecting and interfering two linearly polarized 532-nm laser beams at the position of the BEC.  The beams propagate in the $(\hat{x},\hat{y})$ plane and intersect at an angle of $\sim$9.4$^{\circ}$, producing a periodic array of interference fringes with a measured period of 3.46 $\pm$ 0.05 $\mu$m, with the error primarily due to imaging system magnification calibration uncertainty.    Each beam has a Gaussian profile with a beam diameter of 0.85 mm.  Because the interference region occurs over a much larger region than that occupied by the BEC, the maximum intensity of the interference fringes varies insignificantly over the size of the BEC for the purposes of our experimental assessment.  With 130 mW of optical power in each beam, the corresponding maximum potential height $V_0$ of the comb is $\approx 5.9\, \mu_0$, given in terms of the BEC chemical potential $\mu_0$ calculated for the harmonically trapped BEC without the comb present.

We examined the efficacy of the vortex comb concept for the removal of vortices in the BEC through the following steps and  numerous runs of similar experiments.  First, for each experimental run, we stirred vortices into the BEC as described above.  We then ramped on the comb linearly in time for 1~s, 2~s, or 4~s, and then immediately ramped the beams back off over the same durations, for \emph{total} ramp times $T_{\rm{tot}}$ of 2~s, 4~s, and 8~s.  Finally, we expanded and imaged the BEC, and counted the number of vortices remaining in the BEC in order to compare with the mean numbers of vortices remaining in the BEC after holding for the same total time but without using the comb.  The primary experimental parameters that we adjusted were the comb ramp-on and ramp-down times (which were identical to each other for all cases) and the height of the comb potential.  We did not vary the periodicity of the OL fringes in our experiment, although variations of this parameter are, in principle, experimentally accessible and were examined in the numerical study described in the following sections.    

\begin{figure}
 \includegraphics[width=0.95\columnwidth]{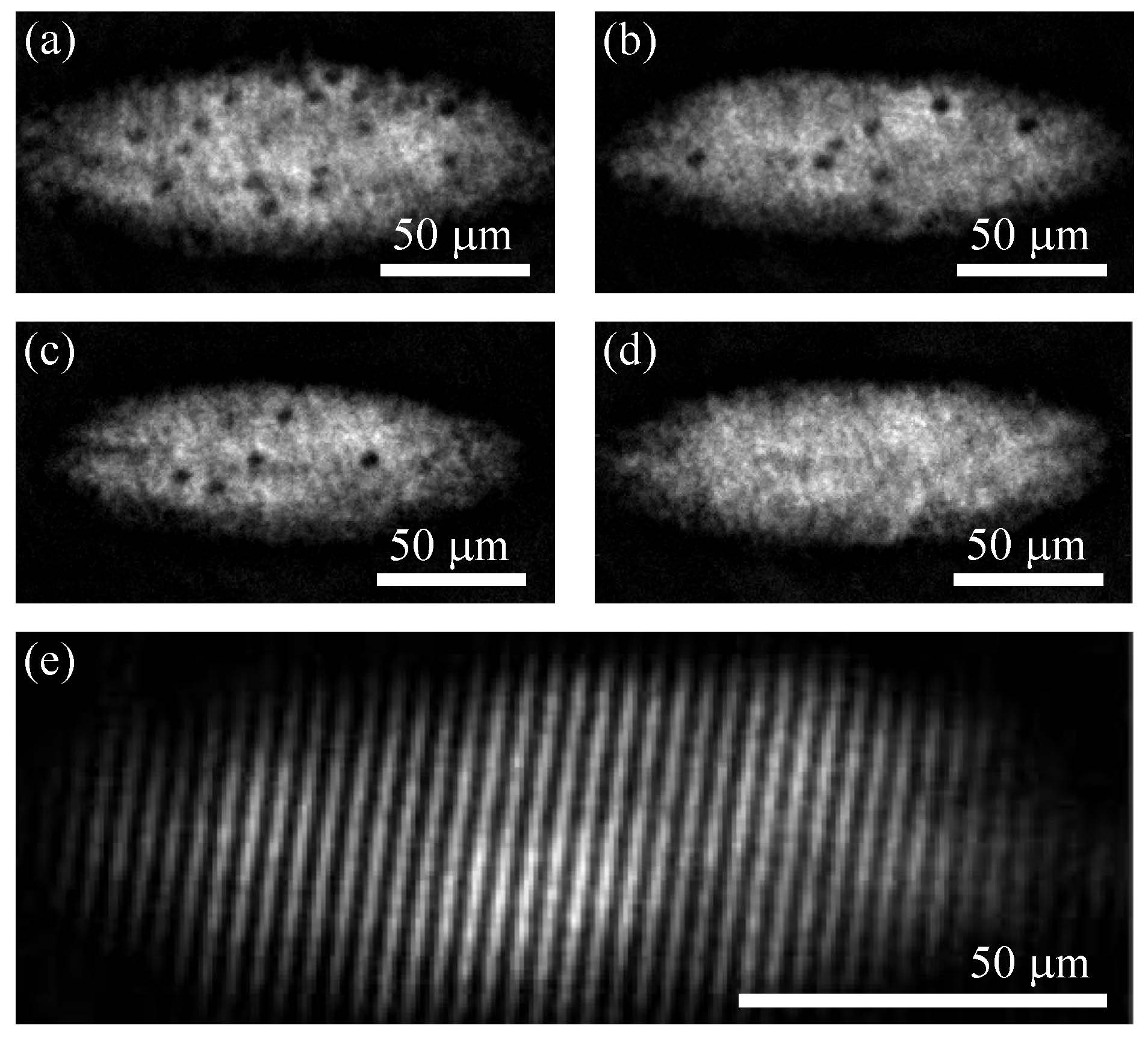}
 \caption{ Representative images of BECs under various conditions:  (a) Immediately after the stirring process, with no hold time, and 12~ms of expansion from the optical trap.  (b) Same as (a) but with 4 s hold time between stirring and expansion.  (c) Same as (b) but with 8 s hold time.  (d) Same as (b) but with a 4-s (total time) ramp of the comb at a peak comb potential depth of $V_0 \sim 5.9\,\mu_0$.  (e) \emph{In situ} image of a BEC with the OL on at a depth corresponding to 14.5 $\mu_0$.  In this case (e), there was no stirring of the BEC prior to the application of the OL.  In all cases, the scale bars indicate the equivalent of a 50-$\mu$m span in object (BEC) space, not on the camera's sensor. %
}
\label{fig:exptimages}
\end{figure}

In Figure \ref{fig:exptimages}(a)--(c), we show representative examples of BECs containing vortices after 0, 4, and 8-second hold times that followed our standard stirring process.  Each case shows a BEC after an additional 12-ms period of expansion after the removal of the optical component of the hybrid trap. Figure~\ref{fig:exptimages}(d) shows an image of a vortex-free BEC after a comb ramp of $T_{\rm{tot}} = 4$ s for $V_0 \sim 5.9 \mu_0$, demonstrating the success of removing vortices from the BEC with these particular comb parameters.  Figure \ref{fig:exptimages}(e) shows an \emph{in situ} and further magnified image of a BEC in the presence of an OL of maximum potential energy $V_0 \sim 14.5 \mu_0$.  In this last case, the BEC was not stirred prior to ramping on the comb.  

\begin{figure}
 \includegraphics[width=1\columnwidth]{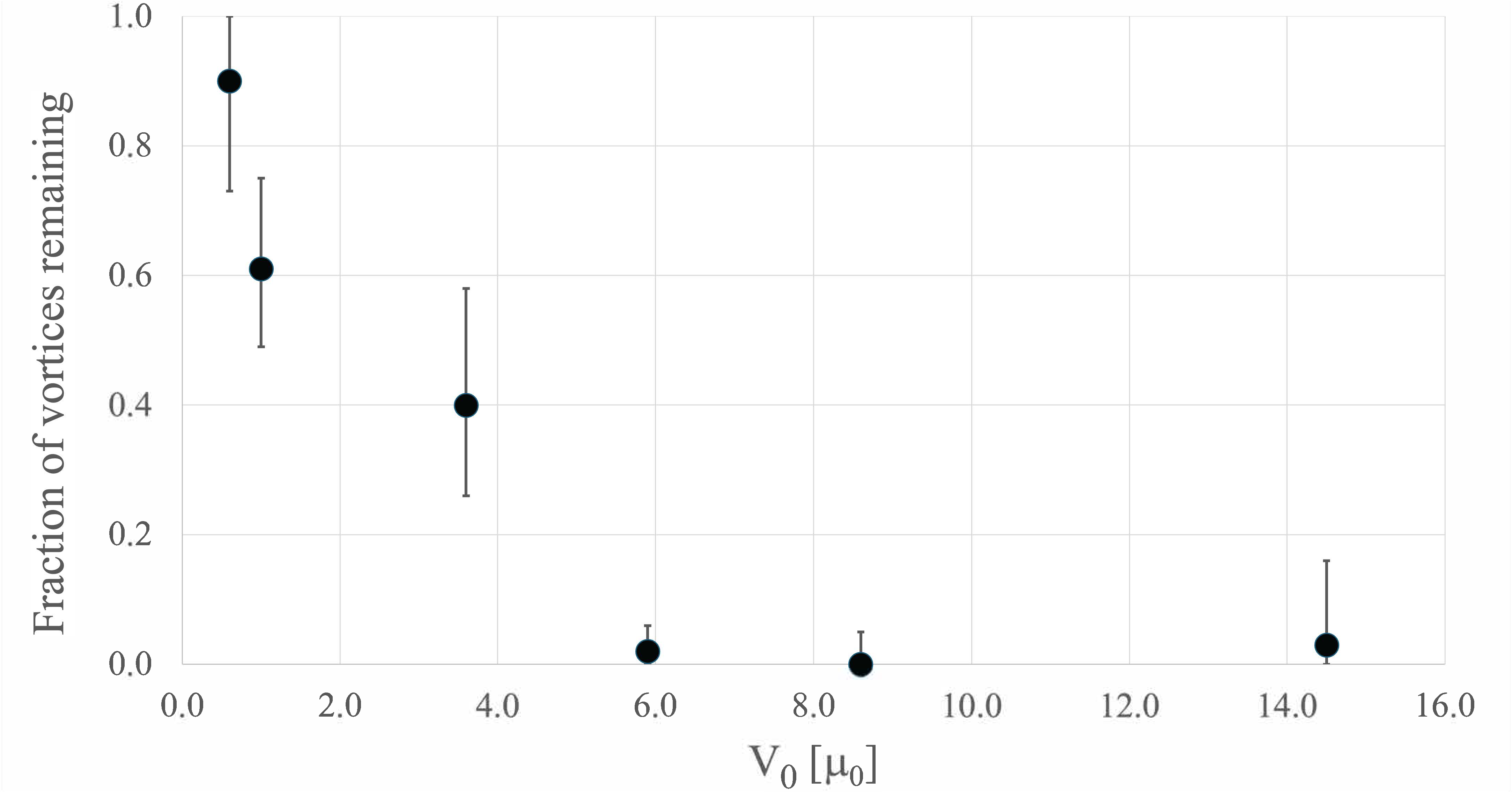}
 \caption{Fraction of vortices remaining after a 4-s (total time) comb ramp, vs.~maximum height $V_0$ of the comb potential in units of chemical potential $\mu_0$. The error bars indicate 95$\%$ confidence intervals, assuming Poisson statistics for the number of vortices generated and the number of vortices removed by the comb. Note that error bars are not necessarily symmetric about the mean fraction, as a fraction of vortices measured must be between 0 and 1.
 }
\label{fig:fractionremaining}
\end{figure}

We first quantified the efficiency of this vortex removal technique by implementing a comb ramp of duration $T_{\rm{tot}} = 4$~s and acquiring data for various values of $V_0$ up to 14.5 $\mu_0$.  For each value of $V_0$, we determined the mean number $N_v$ of vortices remaining after repeated observations for a given set of OL parameters.  Assuming Poisson statistics, we determined a 95$\%$ confidence interval about the observed mean vortex number. Note that such a confidence interval is not necessarily symmetric about the mean.  For example, a mean vortex number arbitrarily close to zero will not have uncertainties that extend to negative values, and may have much larger uncertainties towards larger vortex numbers. We then determined the fraction of vortices remaining by dividing $N_v$ and the associated uncertainty values by 6.8, the mean number of vortices observed after a 4-s hold without the comb applied.  Our results are shown in Fig.~\ref{fig:fractionremaining}.  

Similar to the data of Fig.~\ref{fig:fractionremaining}, we examined the fraction of vortices remaining after applying a combing process with $V_0 \sim 5.9 \mu_0$ for ramps with $T_{\rm{tot}}$ equal to 2 s, 4 s, and 8 s, and we compared the results with the mean number of vortices remaining after the relative matching hold times for comb-free cases.  With $T_{\rm{tot}} = 2$ s, the mean vortex fraction remaining was 0.041(+0.049,-0.027), where the positive uncertainty (towards larger fractions) is reported first within the parentheses, and the negative-trending uncertainty value is second.   For this data set, out of 19 runs of the experiment, vortices were completely removed from 13 runs (68\%).  With $T_{\rm{tot}} = 4$ s, the mean fraction remaining was 0.023(+0.044,-0.019); out of 19 runs of this experiment, vortices were completely removed from 17 runs (89\%).  Note that this result for the $T_{\rm{tot}} = 4$ s case is also indicated in Fig.~\ref{fig:fractionremaining}.  With $T_{\rm{tot}} = 8$ s, the mean fraction remaining was 0.018($+$0.045,$-$0.017); out of 21 runs, vortices were completely removed from 19 runs (90\%).  

From our experimental demonstrations, we draw the following conclusions.  First, the application of the comb is indeed a viable method for vortex removal, as envisioned.  Second, for the time scales of comb application and for the comb parameters studied, there is no observable heating or excitation of the BEC imposed by the comb; the condensate radii are identical (within error limited by shot-to-shot fluctuations) to BEC radii prior to application of the comb.  Third, with $V_0$ approximately in the middle of the range of values examined, the combing efficiency is highly successful irrespective of moderate changes in the timescale of the vortex comb ramp. We therefore conclude that this vortex combing method can be a robust and viable technique that can be applied in experiments where controlled removal of vortices is desired.  

However, there are numerous open questions that our experiments have not addressed.  First, in the experimental procedure, we are unable to specifically pinpoint the physical mechanisms responsible for vortex removal.  Second, with the limited experimental configurations examined ---especially using a fixed OL periodicity ---we are not able to identify whether or not there is an optimum set of comb parameters for vortex removal within a larger parameter space that includes OL periodicity.   Third, we imagine that there are comb parameters that could possibly be too aggressive, i.e., that excite the condensate so much that even if vortices are removed during comb application, new vortices may be generated during the comb removal step.  To address these issues, we turn to an extensive set of numerical simulations for the remainder of this paper.  While the harmonic trap frequencies considered in the numerical studies are chosen to be equivalent to those of the experiment, our numerical studies assume fewer atoms and fully 2D simulations in order to efficiently scan a wide range and various combinations of comb parameters.  We note that while our BECs are indeed three-dimensional, the largest BEC width-to-thickness aspect ratio of $\omega_z/\omega_x \sim 13.8$ suggests that vortices behave as if their dynamics are two-dimensional~\cite{PhysRevA.84.023637}, and that the cores do not significantly bend or tilt out of the $(\hat{x},\hat{y})$ plane.  Therefore, 2D simulations are expected to capture relevant vortex dynamics.  Our primary numerical results that follow support the conclusion that an optimal set of comb parameters does indeed exists for vortex removal, and we identify previously unreported vortex dynamics that may exist in the presence of an OL potential, notably in the case where the lattice spacing is comparable to the the size scale typically associated with a vortex core. 

\section{Theoretical Setup}
\label{sec:theory}

We now turn to the theoretical formulation enabling a systematic analysis of the vortex combing process and accompanying vortex dynamics.
For a fully 3D BEC having the matter-wave field $\psi(x,y,z,t)$, the mean-field dynamics of $\psi$ are analyzed using the  Gross-Pitaevskii equation (GPE)~\cite{Pitaevskii2003,Kevrekidis2015}
\begin{equation} \label{GrossPitaevskiiEquation3D}
i \hbar \frac{\partial \psi}{\partial t} = -\frac{\hbar^2}{2 m} \nabla^2 \psi
 +  \frac{4 \pi \hbar^2 a}{m} |\psi^2| \psi
 + (V - \mu) \,\psi,
\end{equation}
where $\nabla^2$ is the Laplacian (here in 3D), $m$ is the mass of an atom, $\mu$ is the chemical potential, $a$ is the $s$-wave scattering length, and $\psi$ is normalized to the total number of atoms. The external potential $V$ is a combination of a harmonic trapping potential $V_{\mathrm{HT}}$ and a periodic (in $x$) OL potential $V_{\mathrm{OL}}$ of the forms:
\begin{eqnarray}
V(x, y, z,t)&=& V_{\rm HT}(x, y, z) + V_{\rm OL}(x, y, z,t),
\notag 
\\[1.0ex]
V_{\rm HT}(x, y, z) &=& \frac{m}{2}\left(\omega_x^2 x^2 + \omega_y^2 y^2 + \omega_z^2 z^2\right),
\label{eq:V}
\\[1.0ex]
\notag V_{\rm OL}(x, y, z,t) &=& 
f(t) \times\frac{V_0}{2} \sin\left(\frac{2 \pi x}{x_w}\right).
\end{eqnarray}
Here, $\omega_{x,y,z}$ are the angular frequencies associated with harmonic trapping, and $V_0$ and $x_w$ are, respectively, the OL total amplitude (from minimum to maximum potential energy) and fringe spacing (i.e., periodicity).  We henceforth refer to $V_0$ as the ``comb amplitude.'' The dimensionless parameter $f(t)$ regulates the temporal switching (ramping on and off) of the OL (see below), for which $0\leq f(t) \leq 1$.  In this formulation, the comb amplitude $V_0$ plays a role identical to $V_0$ as described in the experimental section above.
This is because although in the experiment the OL potential 
acts through its intensity (i.e., a $\sin^2$ potential), the potential
that we consider in Eq.~(\ref{eq:V}) (i.e., a $\sin$ potential) has
the same shape after an appropriate rescaling of its periodicity and the 
addition of a DC offset (which will not alter the dynamics of the system).

Given the highly oblate nature of the BECs and trapping configurations used in the experiment, we consider for our simulations a quasi-2D configuration for which $\omega_z \gg \omega_x, \omega_y$,  although the experimental values do not strictly fall into this limit.  As noted, this assumption allows efficiency in scanning a large OL parameter set.  We then make the standard ansatz $\psi(x,y,z,t) = \psi_{\text{3D}}(x,y,z,t)\times e^{-i{\omega_z}t/2}$, whereby the wavefunction $\psi_{\rm 3D}$ lies in its ground state along the strong confining ($\hat{z}$) direction. Thus, one may approximate the 3D wavefunction using the decomposition $\psi_{\mathrm{3D}}(x, y, z, t) = \Psi(x, y, t)\times\phi(z)$~\cite{Pitaevskii2003,Kevrekidis2015}, where $\phi(z)$ is the single-particle ground-state wavefunction for the $\hat{z}$ direction, normalized to 1, and $\Psi(x,y,t)$ is normalized to the total number of atoms 
\begin{equation}
    N_{\rm atoms} = \int_{x , y} |\Psi(x, y, t)|^2 \, \mathrm{d}x \, \mathrm{d}y, 
\end{equation}
which, in the current description in the absence
of finite-temperature loss effects, is conserved.
This leads to the effective 2D GPE:
\begin{equation}
\label{EffectiveGPE2D}
i \hbar \frac{\partial \Psi}{\partial t}   
= -\frac{\hbar^2}{2 m}\nabla^2_{x,y} \Psi 
    + g_{\rm 2D} |\Psi|^2 \Psi + (V_{\text{2D}}-\mu)\Psi,
\end{equation} 
where now $\nabla^2_{x,y}$ is the 2D Laplacian in $(x,y)$, the reduced 2D potential is given by
\begin{eqnarray}
    V_{\rm 2D}(x,y,t) &=& V_{\rm HT}(x, y,0) + V_{\rm OL}(x, y,0,t),
    \notag
\end{eqnarray}
and the 2D rescaled (effective) nonlinearity coefficient is given by $g_{\rm 2D}= \sqrt{8\pi}\, a\sigma_z\times\hbar\omega_z$~\cite{Kevrekidis2015}, where $\sigma_z \equiv \sqrt{\frac{\hbar}{m \omega_z}}$ is to be interpreted as a characteristic harmonic oscillator length scale for the $\hat{z}$ direction.  We consider and reference below the experiment's value for $\omega_z$ in order to set an energy and time scale for the simulations that are comparable with those of the experiment. The intrinsic length scale of the condensate, characterizing the approximate width of vortex cores, is given by the healing length $\xi = \sqrt{\frac{\hbar^2}{2 m \mu}}$ for the effective 2D GPE \eqref{EffectiveGPE2D}. 

Finally, to take into account finite-temperature effects, we add a phenomenological dissipation constant $\gamma$ to the 2D GPE, yielding the 2D damped GPE
\begin{equation}
\label{EffectiveGPE2DWithGamma}
(i-\gamma) \hbar \frac{\partial \Psi}{\partial t}   
= -\frac{\hbar^2}{2 m}\nabla^2_{x,y} \Psi 
    + g_{\rm 2D} |\Psi|^2 \Psi + (V_{\text{2D}}-\mu)\Psi.
\end{equation}
The dimensionless parameter $\gamma$  phenomenologically accounts for the interactions of the condensed atoms with a thermal (non-condensed) atomic cloud for temperatures above zero~\cite{PhysRevLett.89.260402,Proukakis_Book,Proukakis_2008,Blakie01092008,ZNG_Book,PhysRevA.69.053623}, and can lead to the time dependence of $N_{\rm{atoms}}$
and thus we now explicitly add this 
time dependence by using the notation $N_{\rm{atoms}}(t)$.
Typical experimentally relevant values of $\gamma$ correspond to $\gamma\sim 10^{-4}$ to $~10^{-3}$~\cite{PhysRevLett.89.260402,Proukakis_Book,Proukakis_2008,Blakie01092008,ZNG_Book,PhysRevA.69.053623}. In what follows we explore the effects of including phenomenological damping on the vortex combing  process.  Unless otherwise noted, our simulations use the following parameters: $m=1.45\times 10^{-25}$ kg and $a = 5.31\times 10^{-9}$ m (as is the case for $^{87}$Rb), $(\omega_x,\omega_y,\omega_z) = 2\pi \times (2.8,8.5,38.5)$ Hz (per the experiment reported above), $\mu = 5.28 \, \hbar \omega_z$, $N_{\rm atoms}(0) = 3.5 \times 10^{5}$, and $\xi = 0.591 \, \mu$m. 
In our simulations below, we vary $\gamma$ over  the range $0 \leq \gamma \leq 0.05$
and, when not explicitly indicated otherwise, we use the prototypical value
$\gamma=0.0005$.

As in the experiments of Sec.~\ref{sec:experiments}, to generate initial conditions for our studies of the vortex comb, we introduce vortices into the BEC prior to switching on the vortex comb. As per the experiments, vortices are formed within the BEC by using an additional time-dependent repulsive barrier that mimics stirring the BEC with a laser beam. The stirring is controlled by the maximum potential energy and shape of the beam and the specific trajectory of the stirring within the BEC.  For fast enough stirring speeds, relative to the local speed of sound, and in the absence of significant acoustic excitation ~\cite{reeves2012classical,wilson2013experimental}, vortex pairs will be produced in the wake of the moving  beam ~\cite{frisch1992transition,winiecki1999pressure,Nee2010.PRL104.160401}. In a manner similar to the experimental setup, we stir using an elliptical orbit at constant speed that follows a constant BEC density contour.  The strength of the stirring beam is linearly ramped up from zero, then kept at a constant strength for an adjustable period of time,  and finally ramped down.   However, the precise protocol to stir and create vortices is not crucial to the main thrust of this work, as we are focusing instead on the subsequent mechanism of vortex removal.

\begin{figure}[H]
\centering
 \includegraphics[width=0.7\columnwidth]{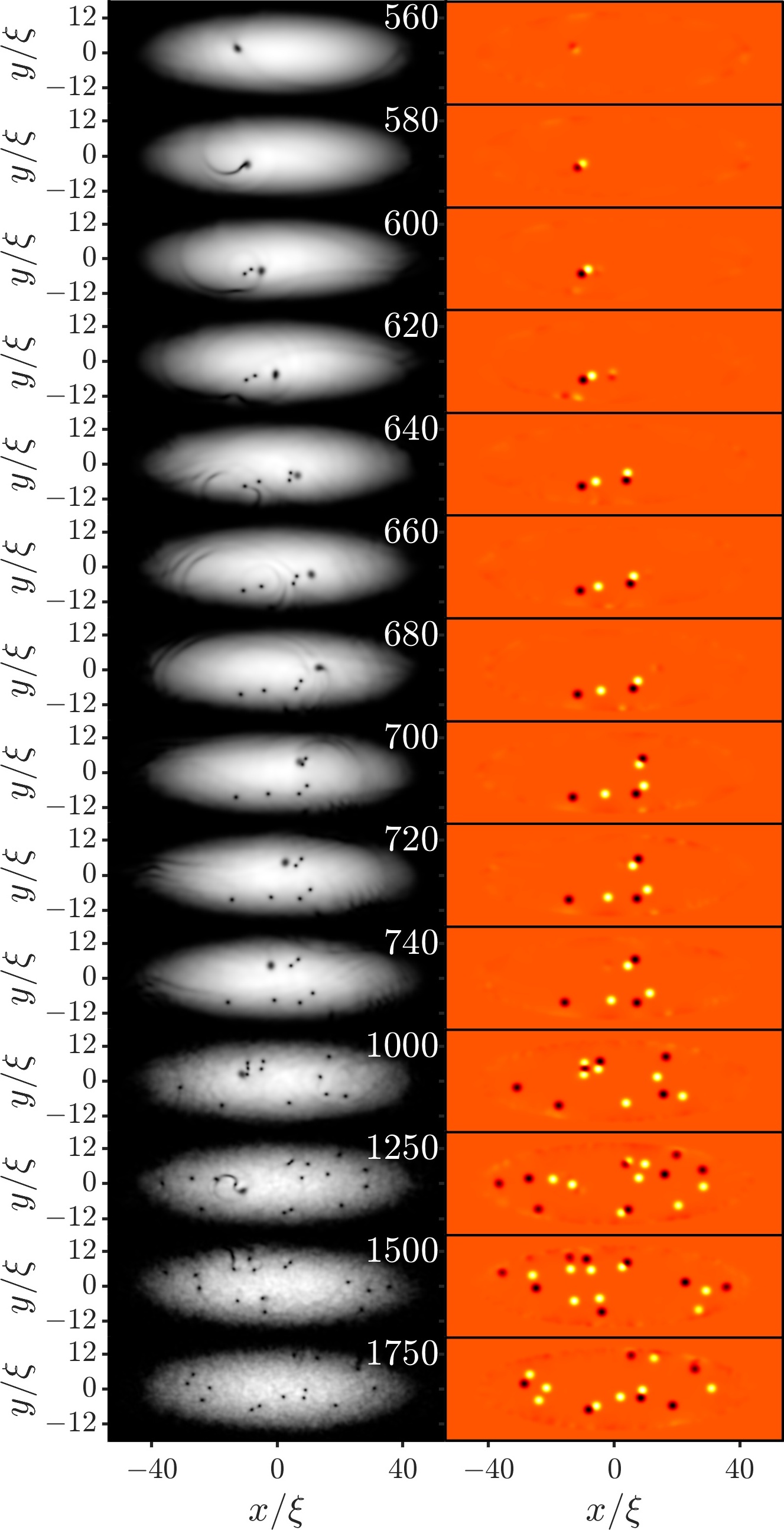}
\caption{(Color online)
Evolution of a typical stirring process nucleating vortices in the BEC. In this case $N_+=8$ positively charged and $N_-=6$ negatively charged vortices remain at the conclusion of the stirring process. The left column of panels depicts the BEC density, with brighter regions corresponding to higher density, and with localized dark regions within the BEC indicating locations of vortices and the stirring potential.  Times of the stirring process (in~ms) are given in the upper right corners of the frames.  The right column of panels depicts the corresponding vorticity profiles (a dark/bright spot corresponds to a negatively/positively charged vortex) at the times indicated.  The system parameters correspond to chemical potential $\mu = 5.28 \, \hbar \omega_z$ and no phenomenological damping ($\gamma=0$).
}
\label{fig:stirring}
\end{figure}

We denote by $N_+$ ($N_-$) the number of vortices with positive (negative) quantized circulation, or charge, about $\hat{z}$
that remain immediately after the stirring process ends, and before the combing begins.  Although the stirring beam creates vortices in pairs (i.e., one with positive and one with negative circulation), not all vortices created by stirring survive through the entire stirring process due to annihilation and damping mechanisms. The number of vortices remaining after stirring is therefore typically smaller than the total number of nucleated vortices.  Additionally, since the stirrer introduces angular momentum into the system, the number of positively and negatively charged vortices remaining after stirring need not be balanced, as some individual vortices may be dissipated at the edge of the BEC. Figure~\ref{fig:stirring} depicts a typical scenario where the stirring process leaves behind $N_+=8$ and $N_-=6$ vortices. The left column depicts the density at the different times using a black-to-white color scheme where white corresponds to the maximum density $\rho_{\rm max}$ computed for all times.  The right column depicts the vorticity using a dark-to-bright color scheme from $w_{\rm min}$ to $w_{\rm max}$ where $w_{\rm min}=-w_{\rm max}$ and $w_{\rm max}$ is the absolute value of the maximum vorticity computed for all times.  The vorticity is computed as the curl of the superfluid velocity which, in turn, is obtained by taking the gradient of the wavefunction's phase~\cite{Kevrekidis2015}.

\section{Numerical Procedure and Results}
\label{Sec:numerics}

\subsection{Description of the numerical experiment}

For the numerical computations that follow, Eq.~\eqref{EffectiveGPE2D} has been discretized (as concerns the Laplacian) using a second-order centered finite difference scheme with homogeneous Dirichlet boundary conditions. We used a standard fourth-order Runge-Kutta integration technique for propagating $\Psi(x,y,t)$ in time according to the damped 2D GPE of Eq.~\eqref{EffectiveGPE2D}.

\begin{figure}
    \includegraphics[width=0.85\columnwidth]{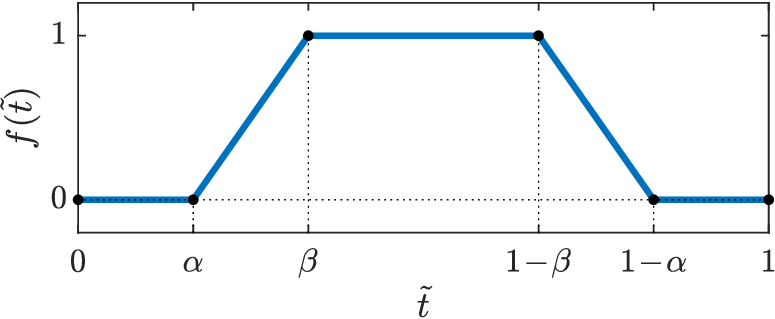}
    \caption{Time profile controlling the amplitude of the
    optical lattice potential. 
    }
    \label{FigHoldProfile}
\end{figure}

\begin{figure}
\centering
\includegraphics[height=11.2cm]{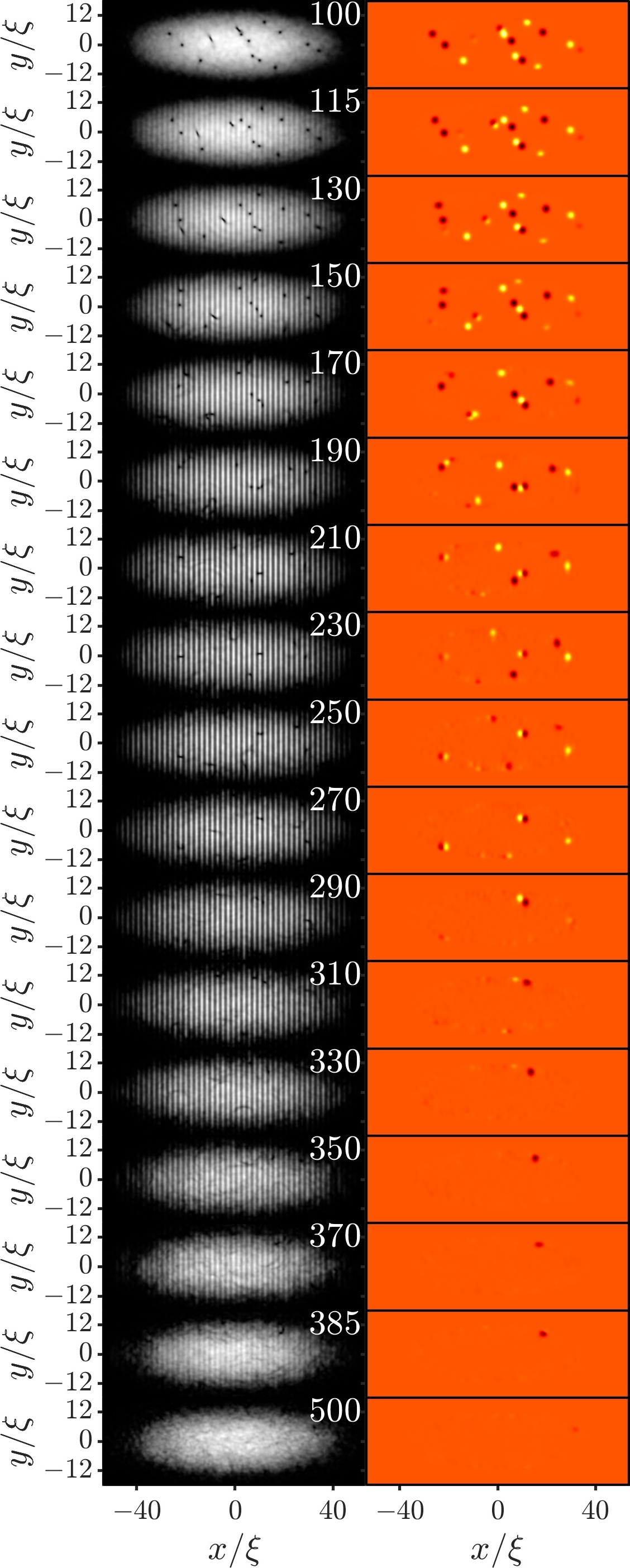}
\,
\includegraphics[height=11.2cm]{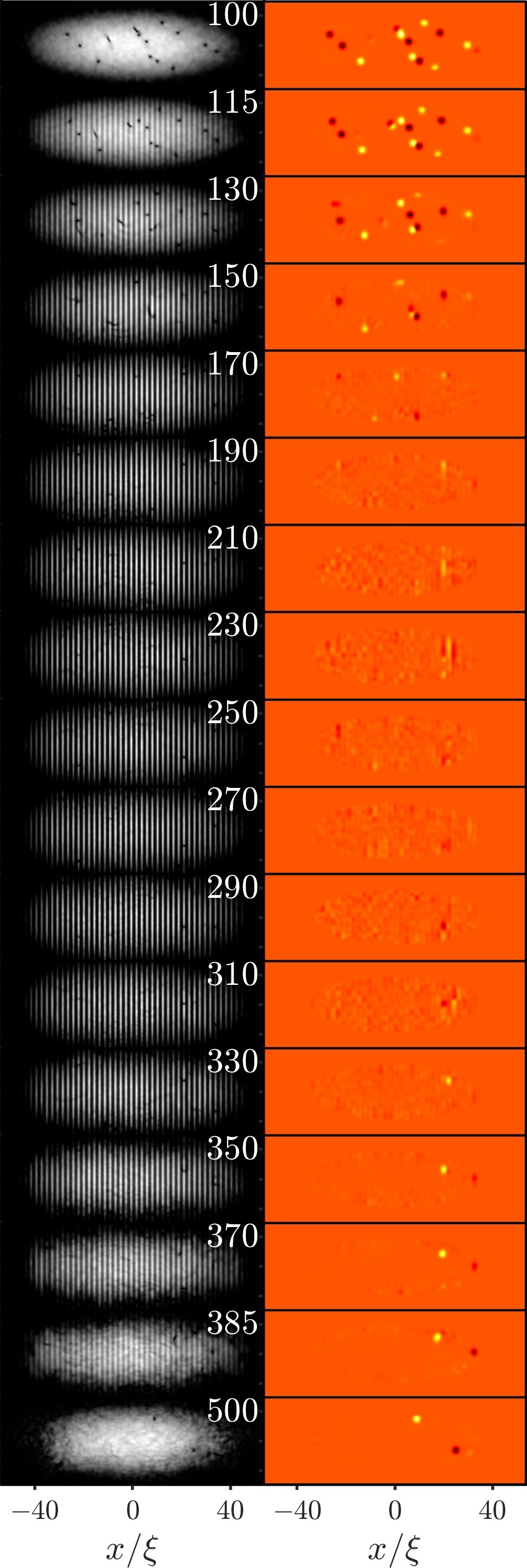}
\caption{(Color online)
Vortex combing for a relatively narrow fringe spacing $x_w=3.5\xi$ 
and maximum comb height of $V_0=1.3\mu$ (left two panels) and $V_0=3.6\mu$ (right two panels).  BEC density and vorticity are shown as in Fig.~\ref{fig:stirring}.  The times (in ms) indicated are with respect to $t=0$, the point at which the combing time profile $f(t)$ is initiated.  For these simulations, we used $\gamma=0$, $T_{\rm tot} = 500$~ms, and ramping parameters $\alpha=1/5$ and $\beta=2/5$.}
    \label{FigDensVortCase12}
\end{figure}

To study the effects of the vortex comb process numerically, we linearly ramp up the OL potential, leave it on for an interval of time, 
and then linearly ramp it down again. The time-dependent amplitude of OL is controlled by the time-dependent profile
\begin{multline}
\label{eq:tent}
    ~f({t})
    =  \begin{cases}
        ~0 & 0 < \tilde{t} < \alpha \\[1ex]
        ~\displaystyle\frac{\tilde{t} - \alpha}{\beta - \alpha} & \alpha  \leq \tilde{t} \leq \beta   \\[1.5ex]
        ~1 & \beta  < \tilde{t} \leq 1 \!-\! \beta \\[1ex]
        ~\displaystyle\frac{1 - \alpha  - \tilde{t}}{\beta  - \alpha} \quad& 1 \!-\! \beta < \tilde{t} \leq 1 \!-\! \alpha  \\[1ex]
        ~0 & 1 \!- \!\alpha  < \tilde{t}\ \leq  1 \\
    \end{cases},\qquad
\end{multline}
where $\tilde{t} \equiv t/T_{\rm tot}$, $T_{\rm tot}$ is the total ramping time, and the dimensionless parameter $\alpha$ controls the time we allow for the BEC to relax before ramping up the OL while the difference $\beta-\alpha$ controls the time we allow for the ramp-up and ramp-down of the OL.  At the end of the ramping process, we also leave a time interval of duration $\alpha T_{\rm tot}$ such that our studies that involve changes to $f(t)$ occur over identical total time intervals $T_{\rm tot}$. A pictorial representation of the relevant ramp-up and -down procedure is depicted in Fig.~\ref{FigHoldProfile}.  This ramp protocol is used in all of our numerical simulations.  Note also that $T_{\rm tot}$ plays the same role here as it does in our experiments, although here the additional parameters of $\alpha$ and $\beta$ allow for further flexibility in holding the OL at its maximum potential energy prior to ramping the OL back down.

\begin{figure}
\centering
\includegraphics[height=11.1cm]{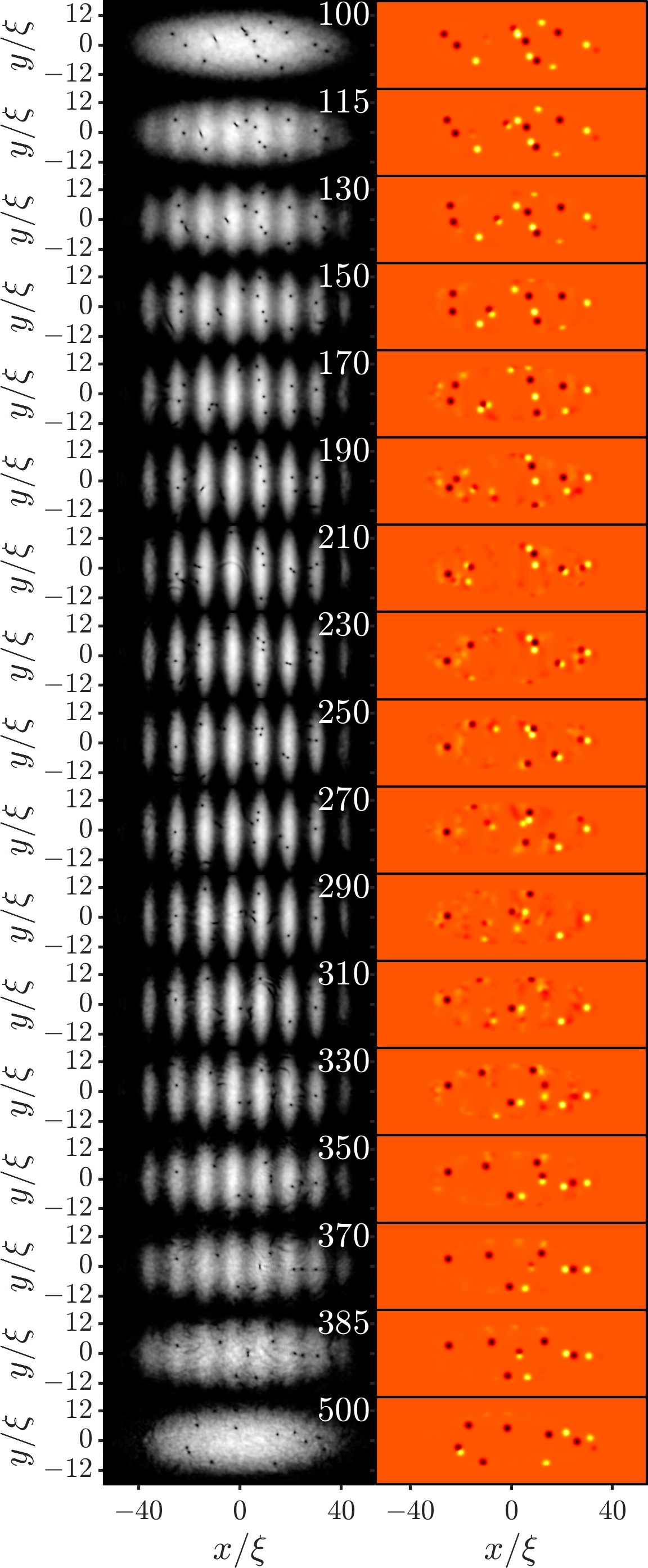}
\,
\includegraphics[height=11.1cm]{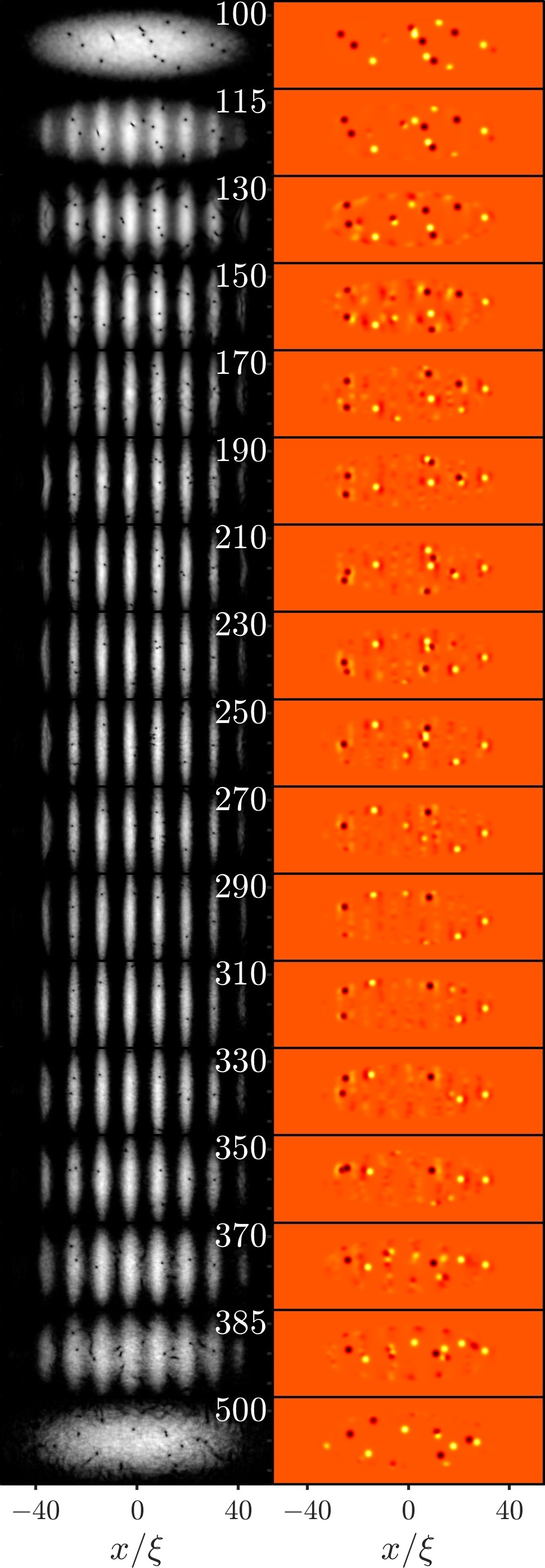}
\caption{(Color online)
Same as in Fig.~\ref{FigDensVortCase12} but for a  wider fringe spacing of $x_w=18\xi$. 
}
    \label{FigDensVortCase34}
\end{figure}

The aims of our simulations are to determine vortex removal efficiency due to the combing process, and to understand the associated vortex dynamics by monitoring the number and appearance of vortices within the BEC during and after the combing process.  As described below in Sec.~\ref{sec:sub:mechanisms}, we undertake a detailed study of the combing process by varying the comb's parameters.  To give context to the detailed results that will be described, we first preview here prototypical examples that show at a large spatial scale the effects of the combing process for two different maximal comb amplitudes $V_0$, first for $x_w = 3.5\xi$ in Fig.~\ref{FigDensVortCase12}, and then for $x_w = 18\xi$ in Fig.~\ref{FigDensVortCase34}.  In all of the cases of these two figures, we chose $T_{\rm tot}=500$~ms, $\alpha=1/5$, and $\beta=2/5$ such that all  five stages of the ramping process (see Fig.~\ref{FigHoldProfile}) last $100$~ms each.

From these large-scale results, it is already apparent that the combing procedure has the ability to significantly reduce or completely eliminate the number of vortices in the BEC, and that a narrow OL periodicity may be more effective at eliminating vortices than wide fringes, whereas a larger comb amplitude $V_0$ is not necessarily more effective than smaller values.

\begin{figure}[htb]
\centering
\includegraphics[width=1.0\columnwidth]{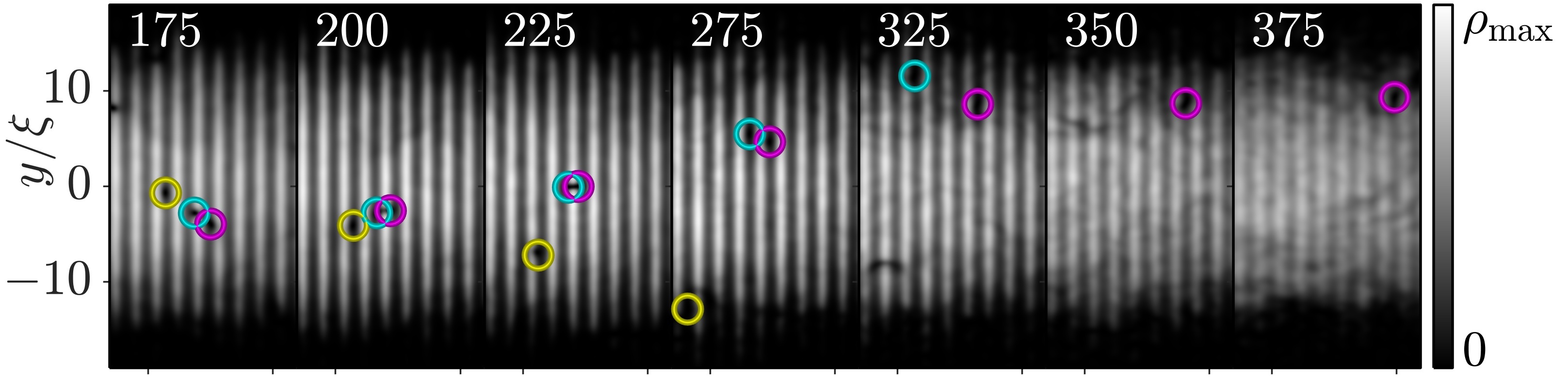}
\includegraphics[width=1.0\columnwidth]{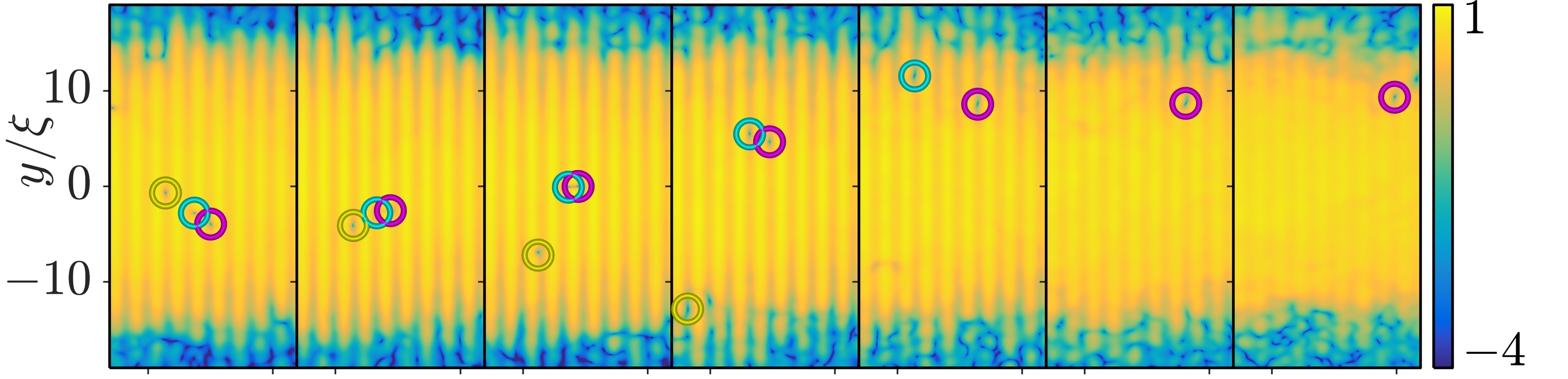}
\includegraphics[width=1.0\columnwidth]{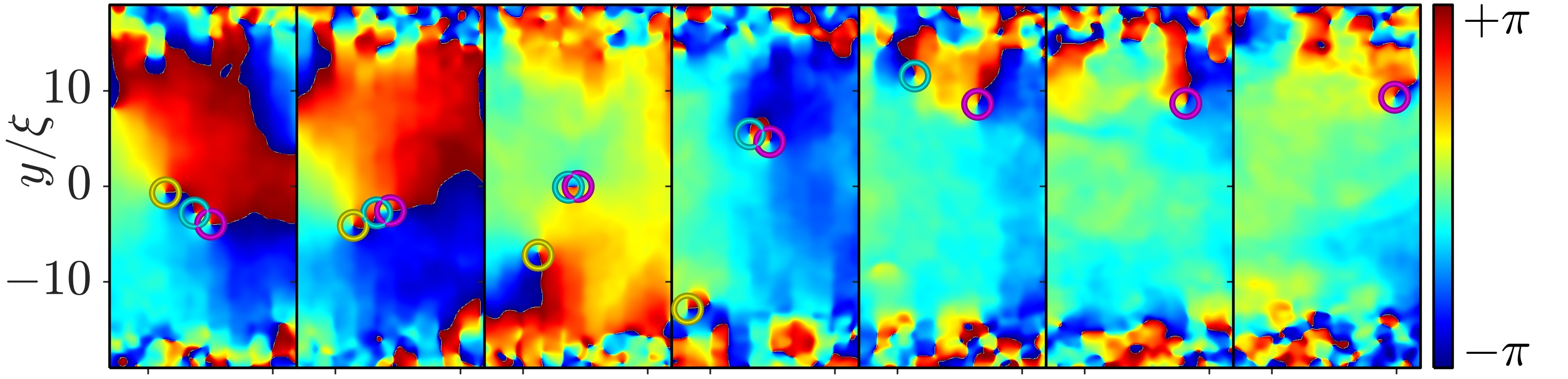}
\includegraphics[width=1.0\columnwidth]{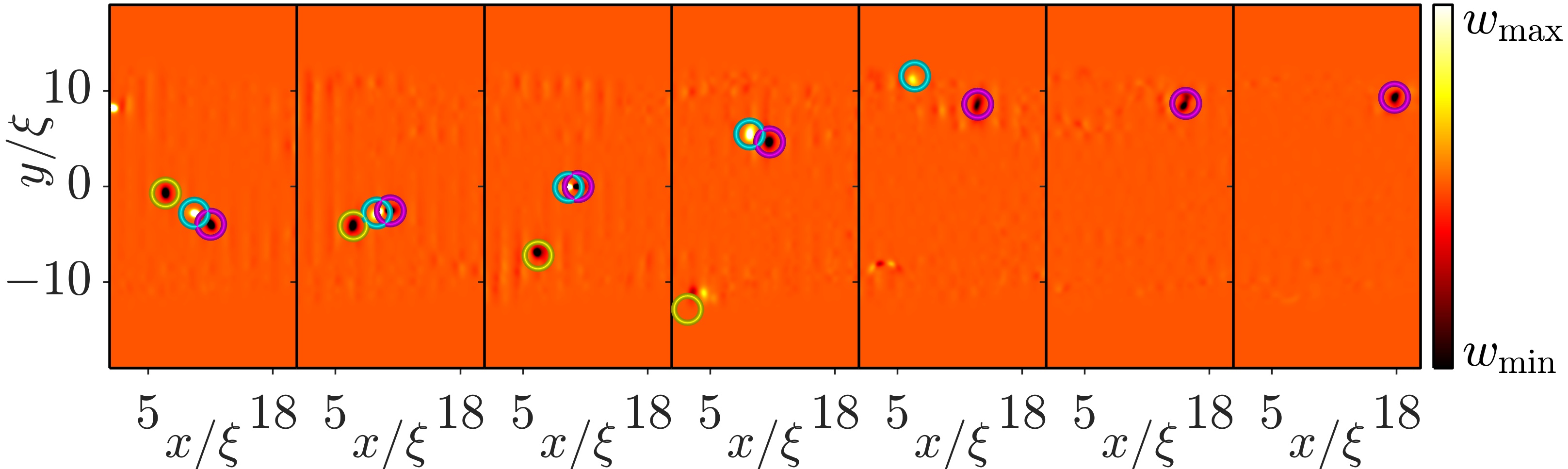}
\caption{(Color online)
Two cases of peripherally channeled vortices (PCVs).  
The four rows of panels correspond, from top to bottom, to density, log (in base ten) of density, phase, and vorticity.  Time (relative to $t=0$) is indicated in~ms in the top left corners of the top row.  Out of the three vortices in the central region of the panel corresponding to $t=175$~ms, the left-most one (see yellow circle) is channeled on its own and gets 
absorbed at the periphery shortly after $t=275$~ms. In contrast, the middle vortex of these three (see cyan circle), a positively charged vortex,  pairs with the right vortex (a negatively charged one; see magenta circle). This vortex pair travels upwards where the vortices separate
as the  positively charged one gets absorbed at the periphery of the cloud after $t=325$~ms, while the negatively charged vortex survives and continues its evolution inside the bulk of the condensate.  The comb parameters correspond to the case $x_w=3.5\xi$ and $V_0=1.3\mu$ (i.e., left set of panels in Fig.~\ref{FigDensVortCase12}). 
The phenomenological damping coefficient is $\gamma = 0.0005$.
\label{Fig_PCV_12}}
\end{figure}

\subsection{Vortex Elimination Mechanisms}
%
In order to better understand the effects of the combing process, we now turn to a systematic description of the vortex dynamics and the mechanisms of vortex elimination that we have observed to occur in the combing procedure. We first list and describe four different mechanisms 
that we have observed and identified as playing a role in reducing the number of vortices during the combing procedure. 
\begin{enumerate}
\item {\em Peripherally channeled vortices (PCV).}
In this case, vortices are channeled along a comb fringe towards the outer edge of the BEC (only when the comb is switched on from 
$\alpha < \tilde{t} < 1 - \alpha$) where they are eventually 
eliminated from the bulk of the BEC. We have identified three different PCV-related scenarios as follows.
(i) A single vortex channeled by a comb fringe that is subsequently absorbed at the outer edge of the BEC cloud; see the vortex identified by the yellow circle in Fig.~\ref{Fig_PCV_12}. Note that when a vortex is located within a dark fringe, where density is non-zero but very low, the density dip of a vortex (that is typical of a vortex in a bulk BEC) is not readily apparent. Therefore, we also show the density on a logarithmic scale to highlight the drop in density that occurs at the phase dislocation.
(ii) A vortex pair that is channeled by a comb fringe and, as the pair approaches the outer edge of the condensate, only one vortex of the pair is eliminated; see the vortices indicated by the cyan and magenta circles in Fig.~\ref{Fig_PCV_12}.
(iii) A vortex pair that is channeled by a comb fringe and both vortices are absorbed at the outer edge of the BEC;  see Fig.~\ref{Fig_PCV_3}.
%

\begin{figure}[htb]
\centering
\includegraphics[width=1.0\columnwidth]{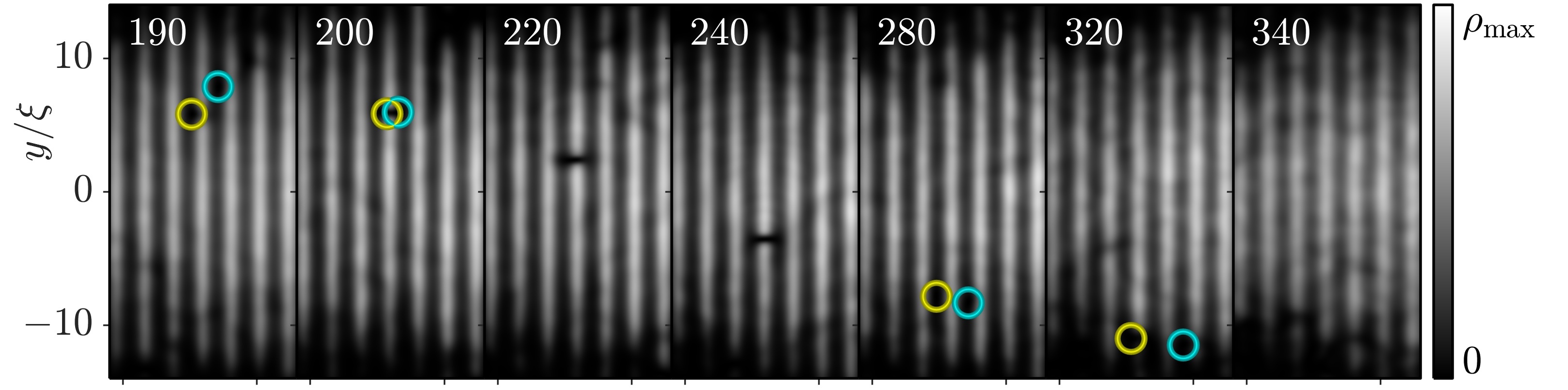}
\includegraphics[width=1.0\columnwidth]{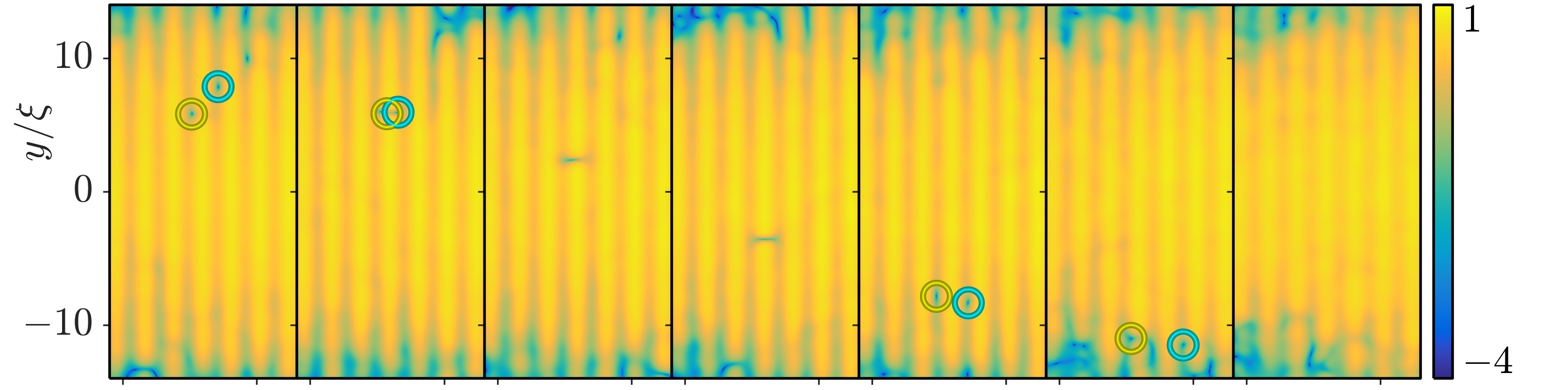}
\includegraphics[width=1.0\columnwidth]{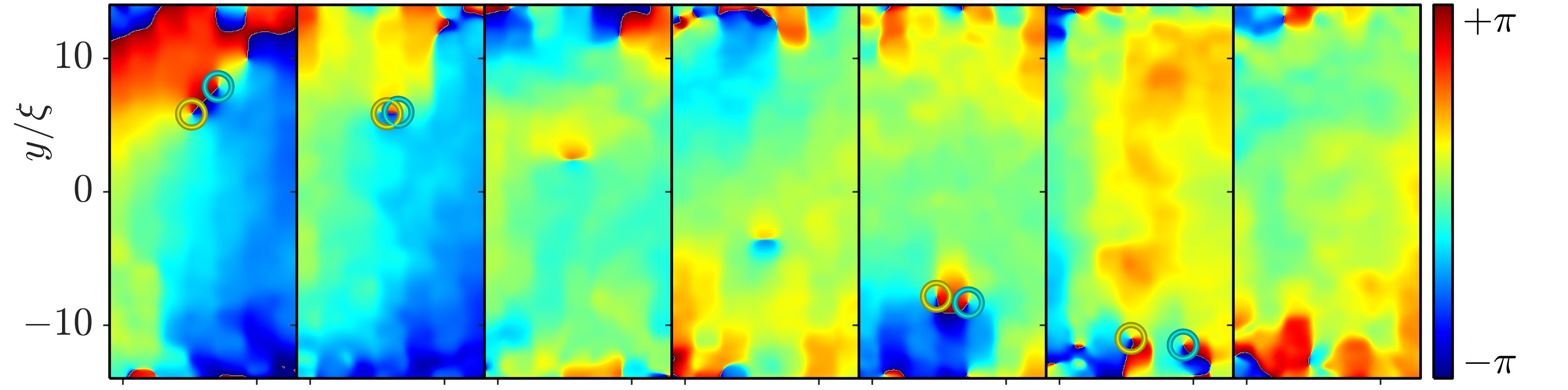}
\includegraphics[width=1.0\columnwidth]{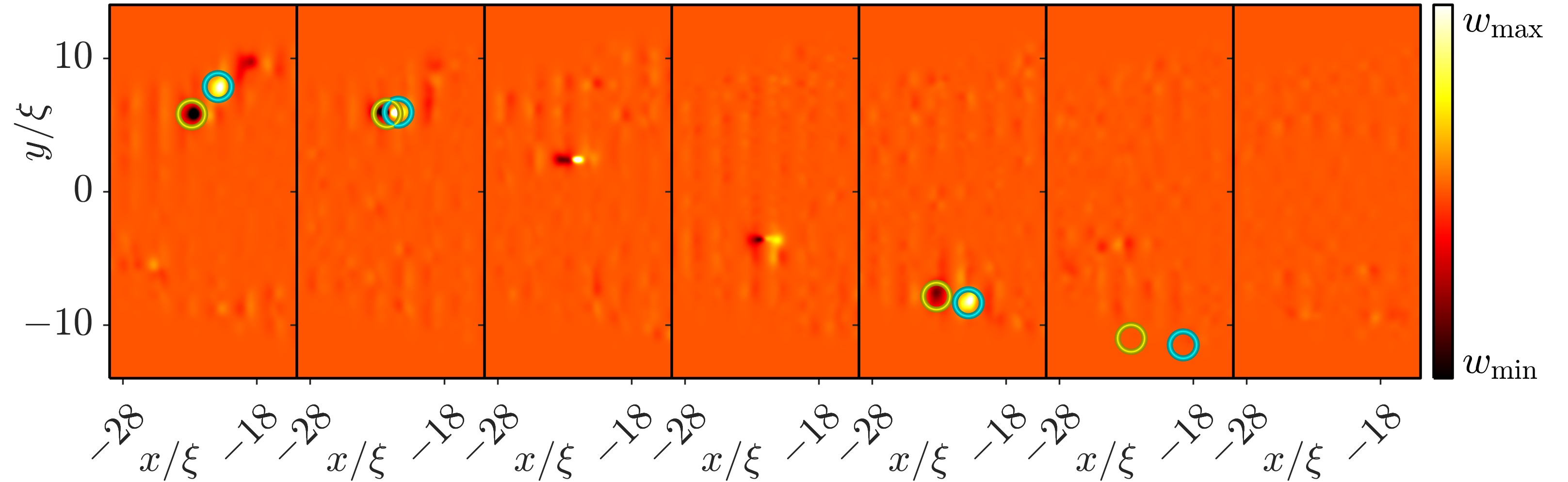}
\caption{(Color online)
A case of a peripherally channeled vortex pair. The items shown in the different panels are the same as in Fig.~\ref{Fig_PCV_12}.  
Two opposite-circulation vortices (see yellow and cyan circles) become bound as a channeled vortex pair around $t=200$~ms. As the pair travels downwards, with the vortex phase singularities located on either side of a bright fringe, the vortices lose their individual 2$\pi$ phase slips and together turn into a JR soliton-like complex. When this complex approaches the bottom edge of the cloud, the phase slips reappear and the vortices start to separate.  Both vortices are eventually absorbed by the periphery of the cloud shortly after $t=320$~ms.  The OL parameters correspond to the case $x_w=3.5\xi$ and $V_0=1.3\mu$ (i.e., left set of panels in Fig.~\ref{FigDensVortCase12}).  Note that the circles depicting the vortex positions are only shown when the corresponding 2$\pi$ phase winding is present.
\label{Fig_PCV_3}}
\end{figure}

\begin{figure}[htb]
\centering
\includegraphics[width=1.0\columnwidth]{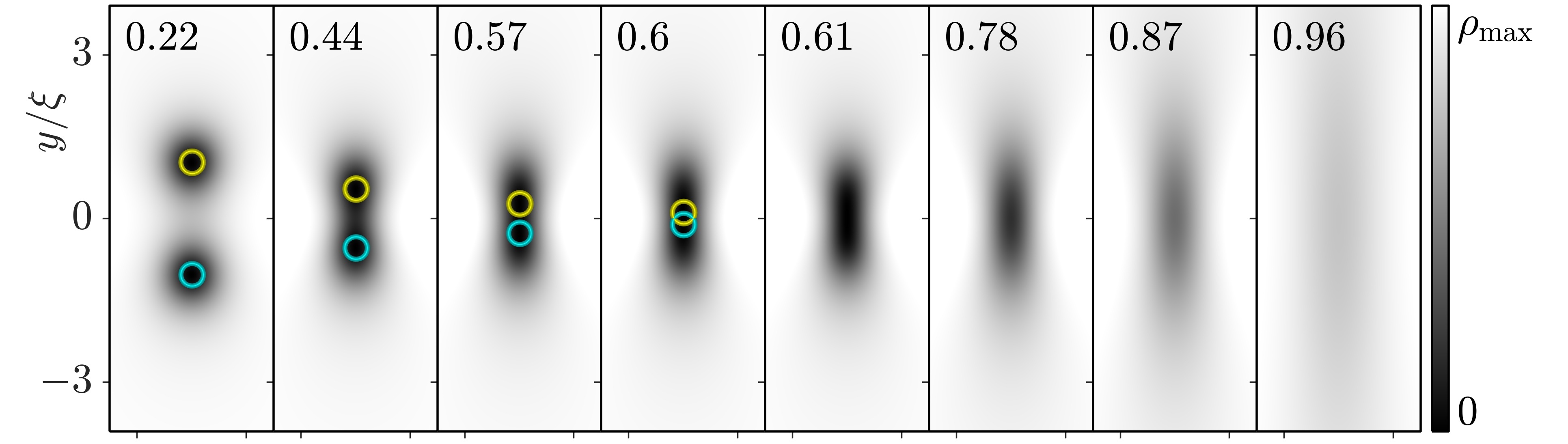}
\includegraphics[width=1.0\columnwidth]{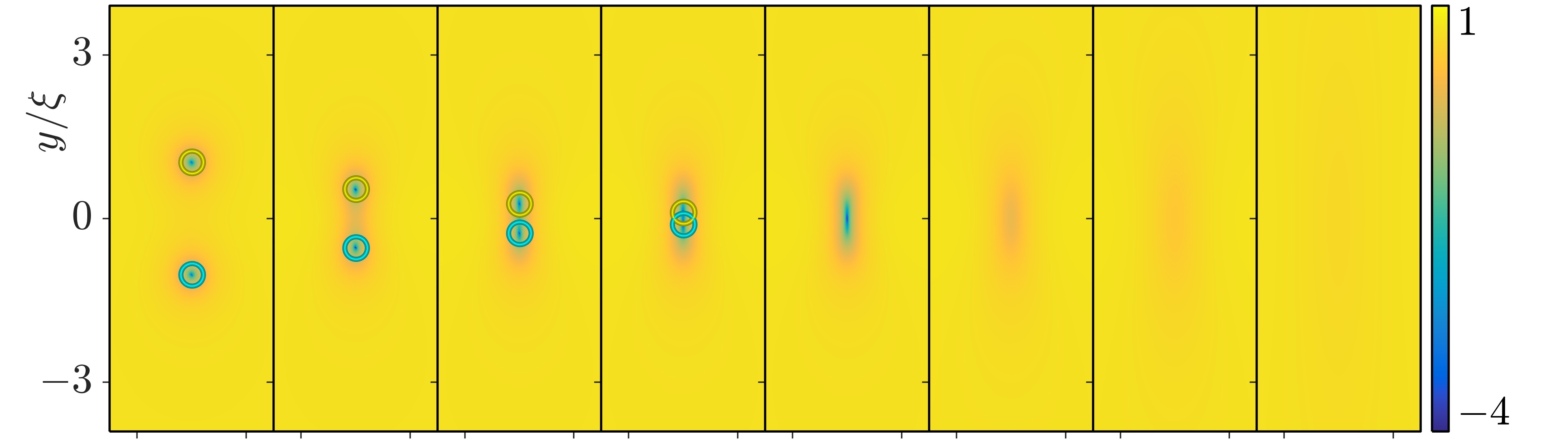}
\includegraphics[width=1.0\columnwidth]{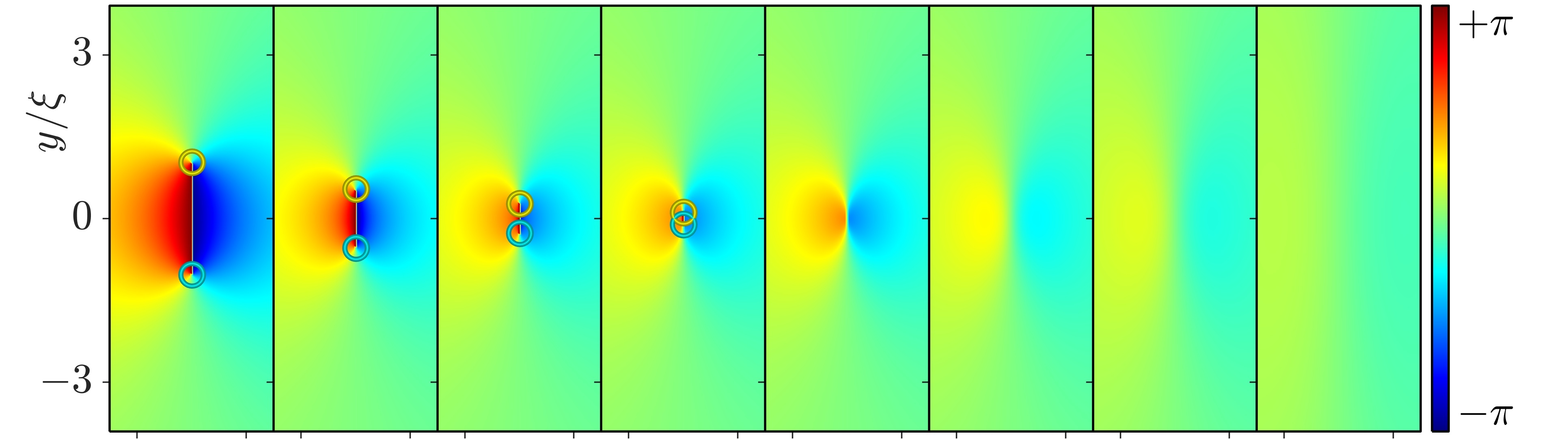}
\includegraphics[width=1.0\columnwidth]{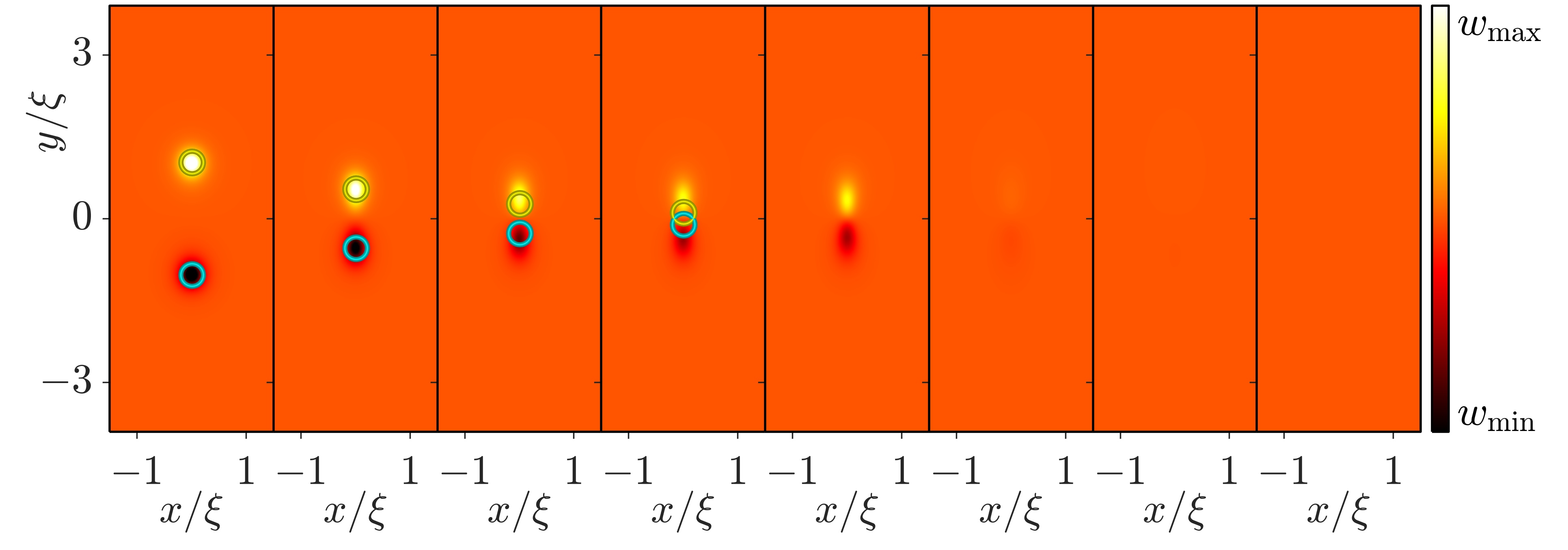}
\caption{(Color online)
Transition between a vortex pair and a JR soliton in an homogeneous background with $\mu=5.28 \, \hbar \omega_z$ as the velocity is increased.
The panels have the same meaning as in previous figures; however, the labels in the top row here indicate the velocity, as a fraction of the speed of sound, of the co-traveling frame where these solutions were obtained. Note that the circles depicting the vortex positions are only shown when the corresponding 2$\pi$ phase winding is present.  
\label{Fig_JR}}
\end{figure}

\begin{figure}[htb]
\centering
\includegraphics[width=1.0\columnwidth]{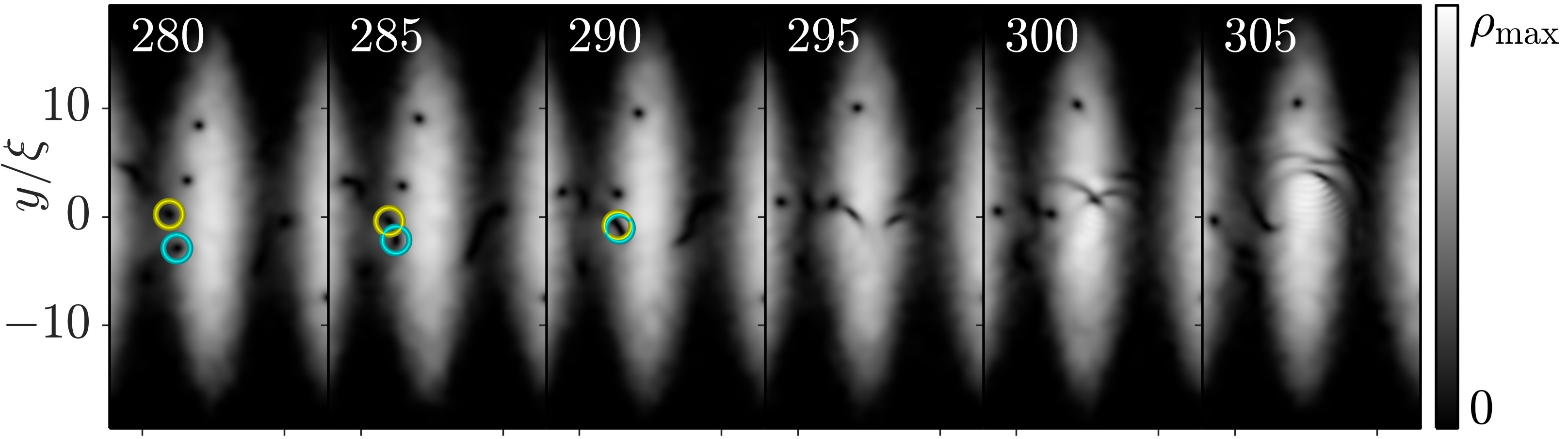}
\includegraphics[width=1.0\columnwidth]{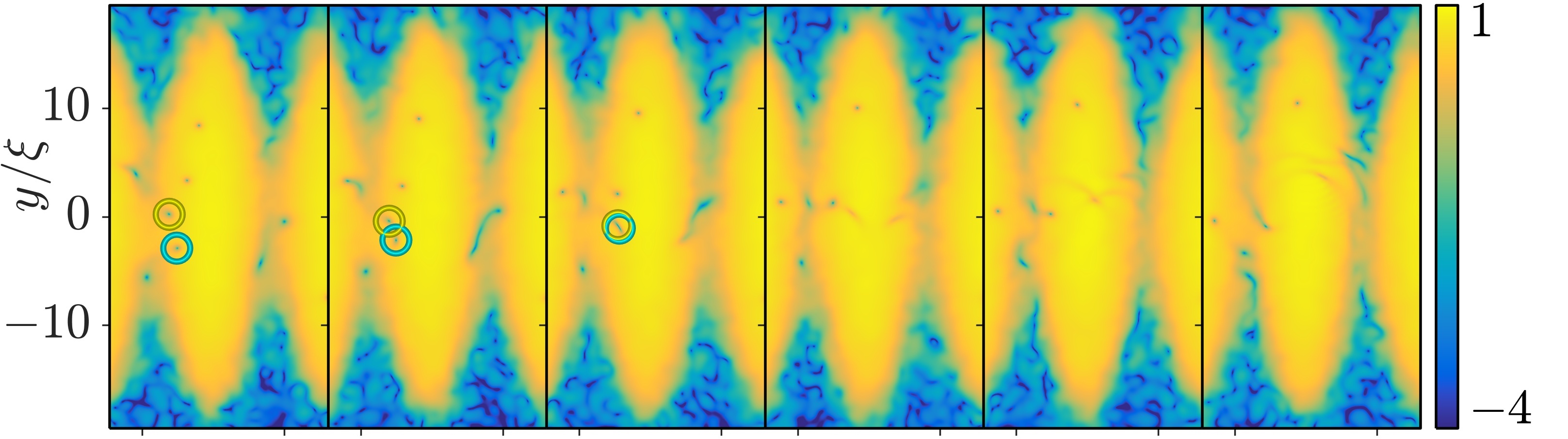}
\includegraphics[width=1.0\columnwidth]{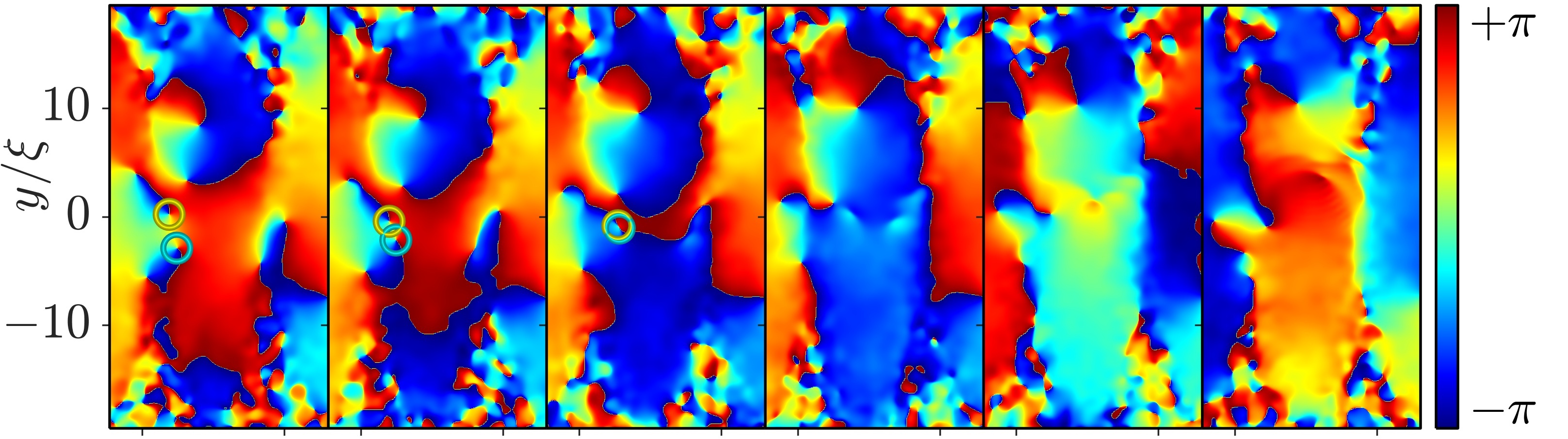}
\includegraphics[width=1.0\columnwidth]{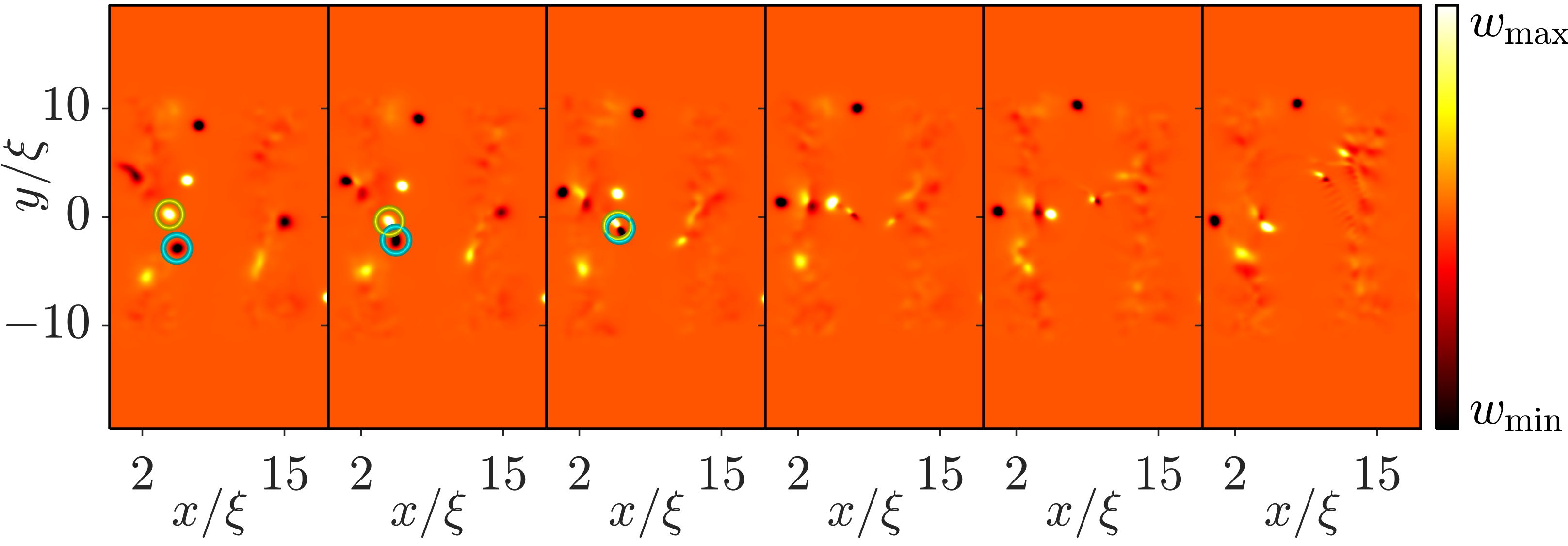}
\caption{(Color online)
Vortex--anti-vortex (V-AV) elimination.
When the BEC fringes are sufficiently wide, V-AV eliminations are observed. In this case, positively (yellow circle) and negatively (cyan circle) charged vortices collide and annihilate each other shortly after $t=290$~ms creating a sound wave that heavily perturbs the remaining vortices inside the fringe.  The comb parameters correspond to the case $x_w=18\xi$  and $V_0=1.3\mu$ (i.e., left pair of panels in Fig.~\ref{FigDensVortCase34}).
}
\label{Fig_AV_case21}
\end{figure}

\begin{figure}[htb]
\centering
\includegraphics[width=1.0\columnwidth]{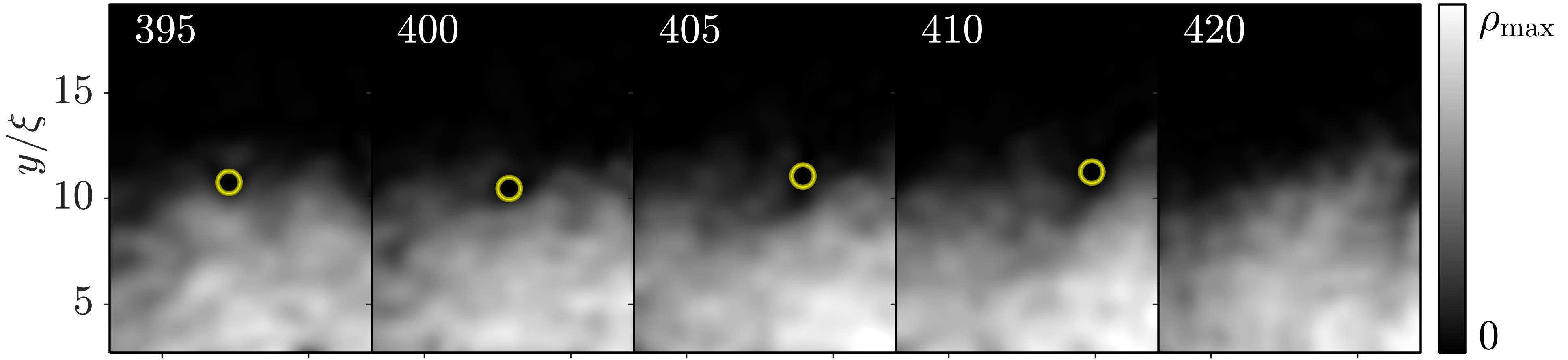}
\includegraphics[width=1.0\columnwidth]{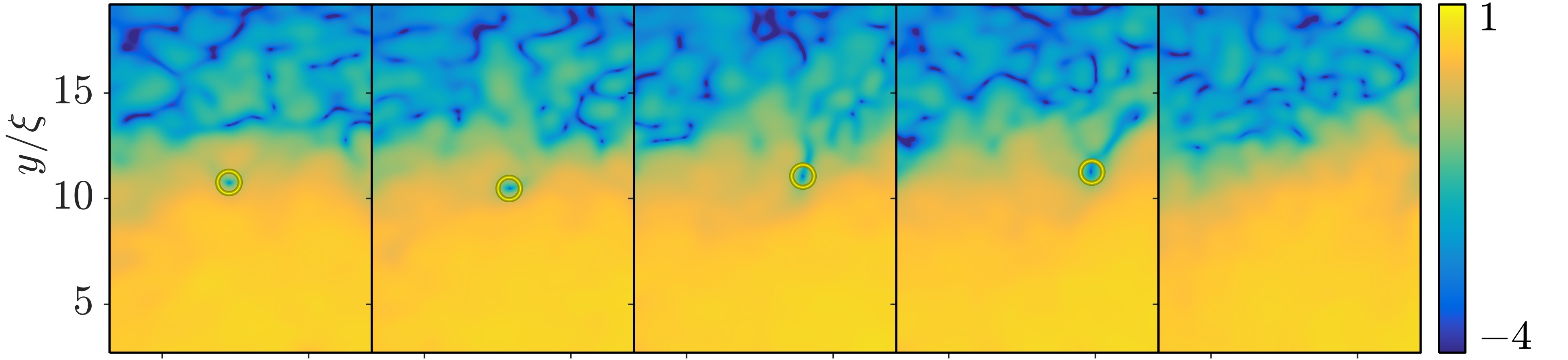}
\includegraphics[width=1.0\columnwidth]{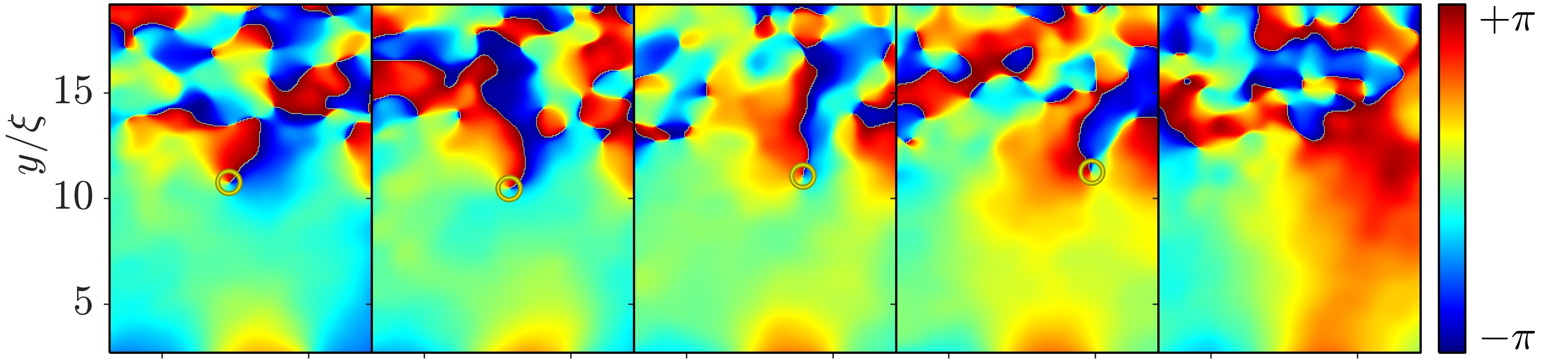}
\includegraphics[width=1.0\columnwidth]{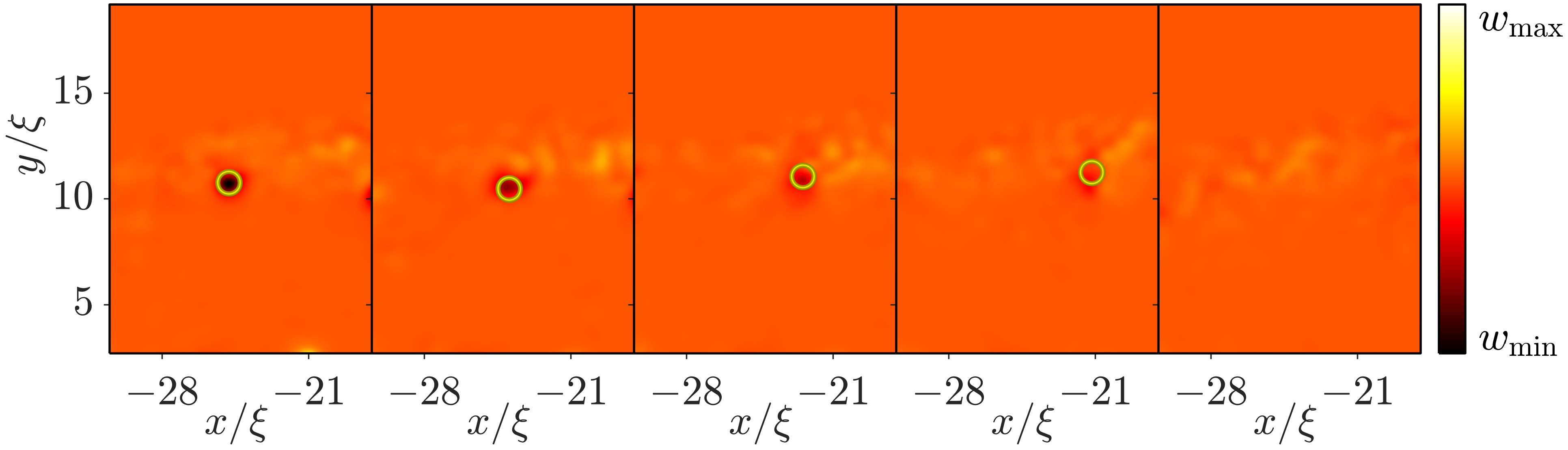}
\caption{(Color online)
Damped vortex (DV) case. 
The comb parameters correspond to $x_w=17.95\xi$ and $V_0=1.4\mu$ (i.e., parameter values close the ones in the left set of panels in Fig.~\ref{FigDensVortCase34}) with a phenomenological damping coefficient of $\gamma=0.0005$. 
}
\label{Fig_DV}
\end{figure}

It is interesting to note that in the last case, as the vortices are channeled through the BEC comb fringe, the phase windings associated with the individual vortices are temporarily lost as the vortex pair morphs into a state reminiscent of a Jones-Roberts (JR) soliton~\cite{CAJones_1982,Jones1986,Berloff2004,bongs,ashtonJR}; see, in particular, the snapshots between $200$ and $240$ ms. In Fig.~\ref{Fig_JR} we show stationary, co-traveling, configurations in free space consisting of a vortex pair (at small velocities) that transforms into a JR soliton (for larger velocities). These configurations were obtained using a standard fixed-point iteration scheme for a stationary state mounted on a co-moving reference frame.  Note that the merger and subsequent separation of the vortex pair in Fig.~\ref{Fig_PCV_3} is tantamount to a transition between a vortex pair and a JR soliton (which lacks the phase singularities of the vortices) as shown in Fig.~\ref{Fig_JR}; see the corresponding phase and vorticity profiles.  

\item {\em Vortex--anti-vortex (V-AV) annihilation.}
This elimination scenario is the commonly envisioned one when two vortices of opposite charge come sufficiently close together that they annihilate each other~\cite{frisch1992transition,Nee2010.PRL104.160401,Kwon2021}. An example of such a mechanism is shown in Fig.~\ref{Fig_AV_case21}. These V-AV annihilations are absent in the case of sufficiently narrow channels, as described below. 

\item {\em Damped vortices (DV).}
In this event type, 
which is only ascribed after the
OL has been turned off (i.e., for $\tilde{t} > 1 - \alpha$), due to the effects of thermal damping at positive temperatures, a vortex near the edge of the BEC will migrate further towards the edge of the BEC where its circulation will be irreversibly dissipated due to the BEC's inability to support flow around the phase 
singularity~\cite{Fedichev1999,Svidzinsky2000,fetter_svidzinsky_2001,PhysRevA.69.053623,Jackson2009,Freilich2010,Rooney2010,Yan2014,yongilshin} (see Fig.~\ref{Fig_DV}). More details on the effects of incorporation thermal damping are given below in Sec.~\ref{sec:sub:optim}.

\item {\em Density-phase-separated vortices (DPSV).}
To the best of our knowledge, the vortex dynamics involved in the vortex elimination mechanism described here have not been previously observed and studied. In this mechanism, the vortex phase singularity separates from the density profile of the vortex core.  More explicitly, the phase singularity disappears into the low-density troughs of the BEC when the comb is present, while the density profile of the vortex morphs into a soliton-like structure associated with a density modulation.  This new excitation is dislocated from the phase singularity now confined within the density trough, and the two features appear to further evolve independently. The density excitation created by this mechanism is reminiscent of a JR soliton, while the detached phase slip migrates to and is eventually ``lost'' in the low-density background at the edge of the condensate. We refer to such an excitation as a ``density-phase-separated vortex'' (DPSV).  An example of this DPSV process is shown in Fig.~\ref{Fig_DPSV}. 

\end{enumerate}

As would be expected, not all of the vortices are eliminated by the combing process for all parameter ranges studied. The vortices that survive in the BEC  after the end of the combing process (or that were created during the combing process itself; see below) are included in our statistical characterizations below, contributing to either a total number of $N_+$ (positively charged) or $N_-$ (negatively charged) vortices remaining after the combing process has been concluded.

\begin{figure}[htb]
\centering
\includegraphics[width=1.0\columnwidth]{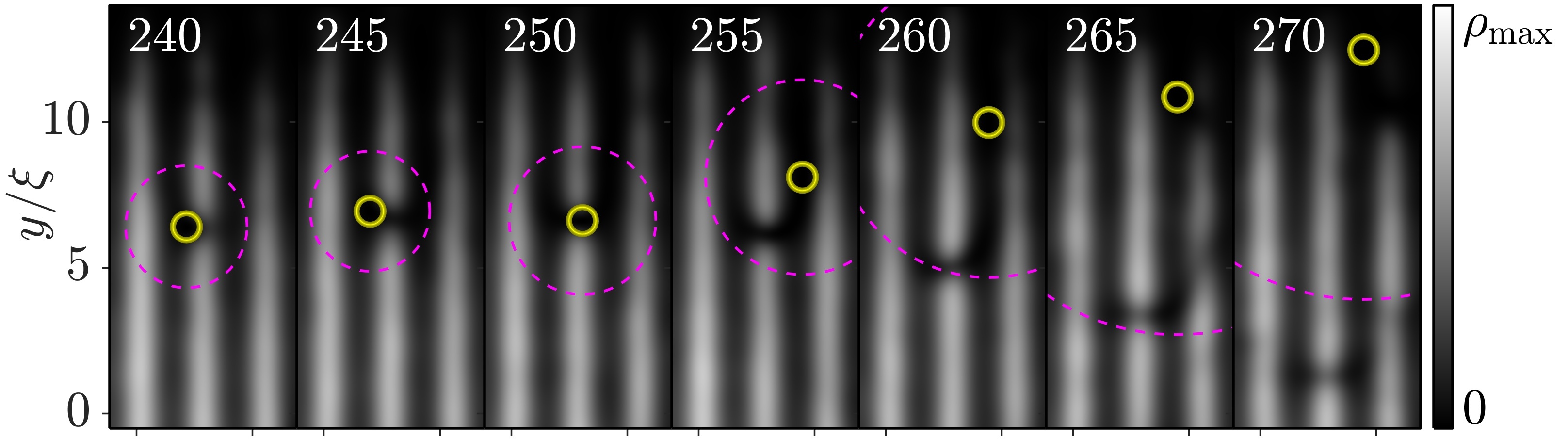}
\includegraphics[width=1.0\columnwidth]{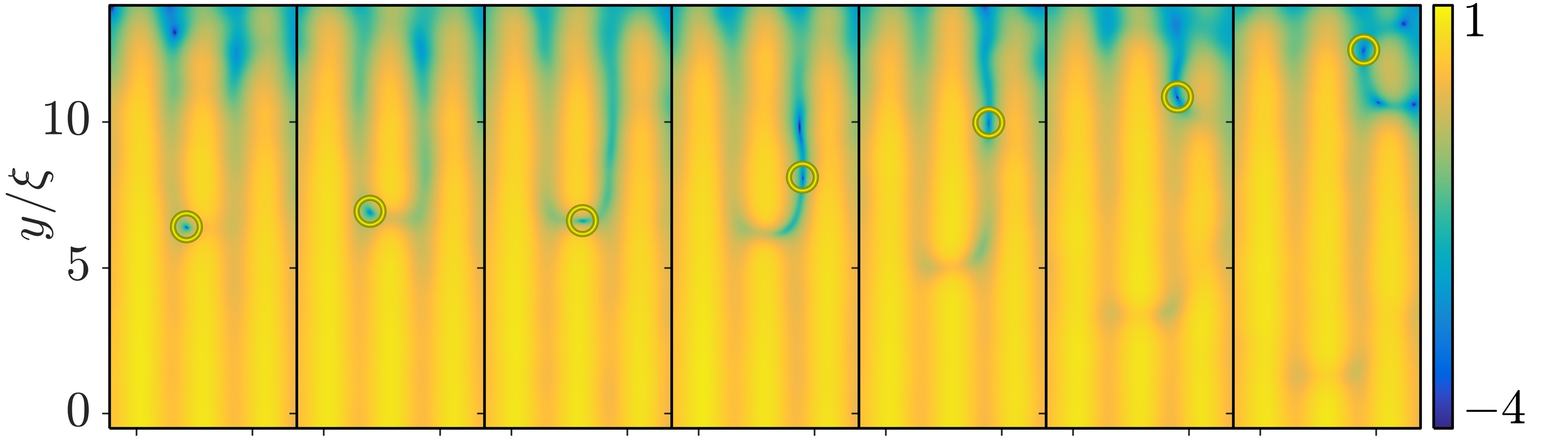}
\includegraphics[width=1.0\columnwidth]{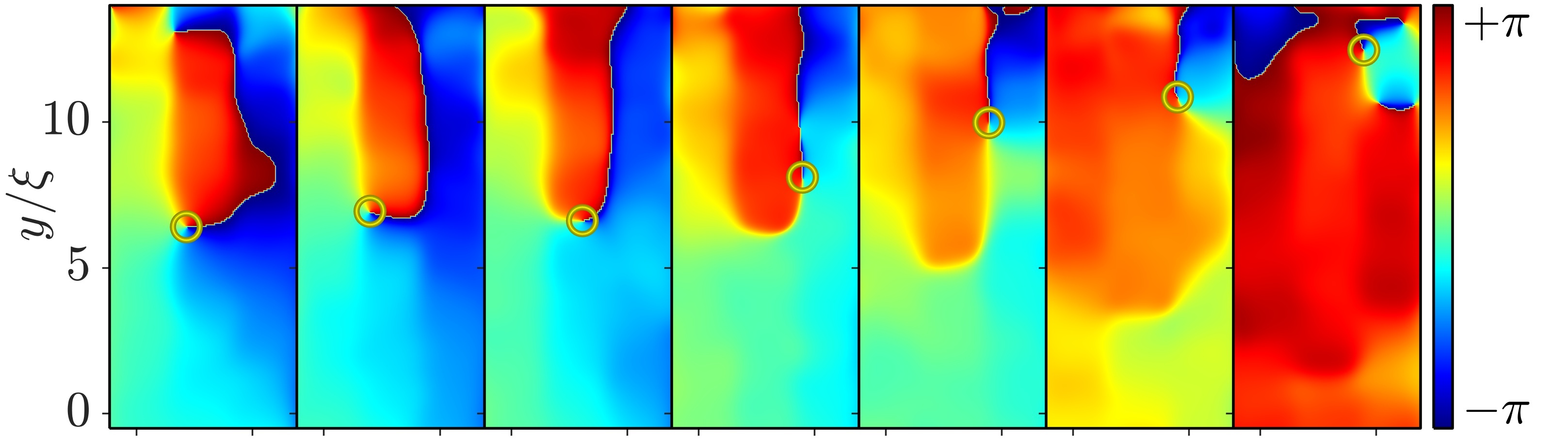}
\includegraphics[width=1.0\columnwidth]{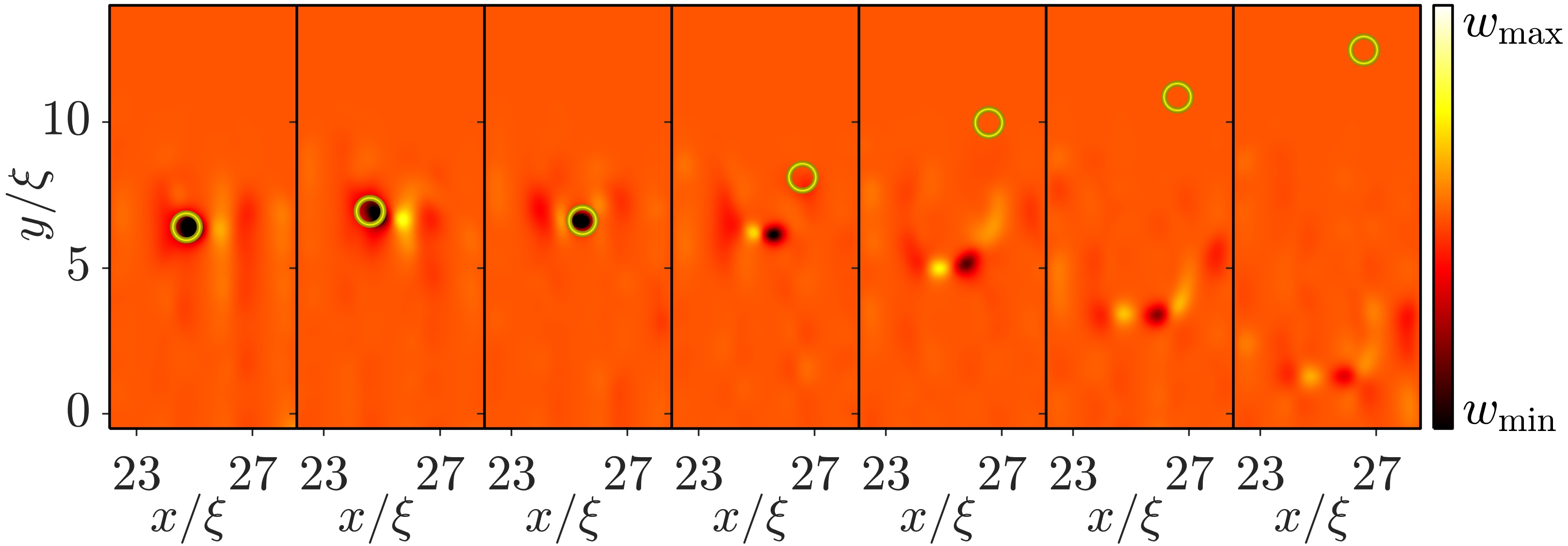}
\caption{(Color online)
Single vortex turning into sound by the DPSV mechanism, for $x_w=3.5\xi$ and $V_0=1.3\mu$ (i.e., left set of panels in Fig.~\ref{FigDensVortCase12}).
The small yellow circle denotes the position of the phase singularity of the vortex, while the larger dashed magenta circle in the top row of density images depicts a circle with a radius equivalent to one quarter of the local healing length calculated at the location of the phase singularity, indicating how large the vortex would be if it existed in a homogeneous background having the low density of the trough. Of primary note in this figure is the separation of the phase singularity indicated by the yellow circles from the density dip observable in the upper two rows of density and log(density) images. The density-phase separation occurs between times of $t=250$~ms and $t=255$~ms. 
}
\label{Fig_DPSV}
\end{figure}

\subsection{Vortex Elimination Mechanisms: 
Observations Regarding $x_w$ and $V_0$ }
\label{sec:sub:mechanisms}

We now turn to a numerical examination of the effects of the choice of comb parameters $x_w$ and $V_0$ on the efficiency of vortex removal due to the general combing protocol described above.  In this subsection, we summarize our general observations obtained from many simulations.  For the bulk of our simulations, we use the choice of trapping parameters and atom number described in Sec.~\ref{sec:theory}, and begin each simulation with precisely the same initial state of the BEC with 18 vortices that is shown in the initial frames of Fig.~\ref{FigDensVortCase12} and Fig.~\ref{FigDensVortCase34}.  

We briefly discuss our general outcomes and observations regarding the impacts of our choices of $V_0$ and $x_{\rm w}$ on the different vortex removal mechanisms previously discussed.  The OL parameters were varied so that we can search for a set of  parameters for which vortex elimination is optimized, as discussed below in Sec.~\ref{sec:sub:optim} alongside our data indicating simulation results.  

\begin{figure}
    \includegraphics[width=0.99\columnwidth]{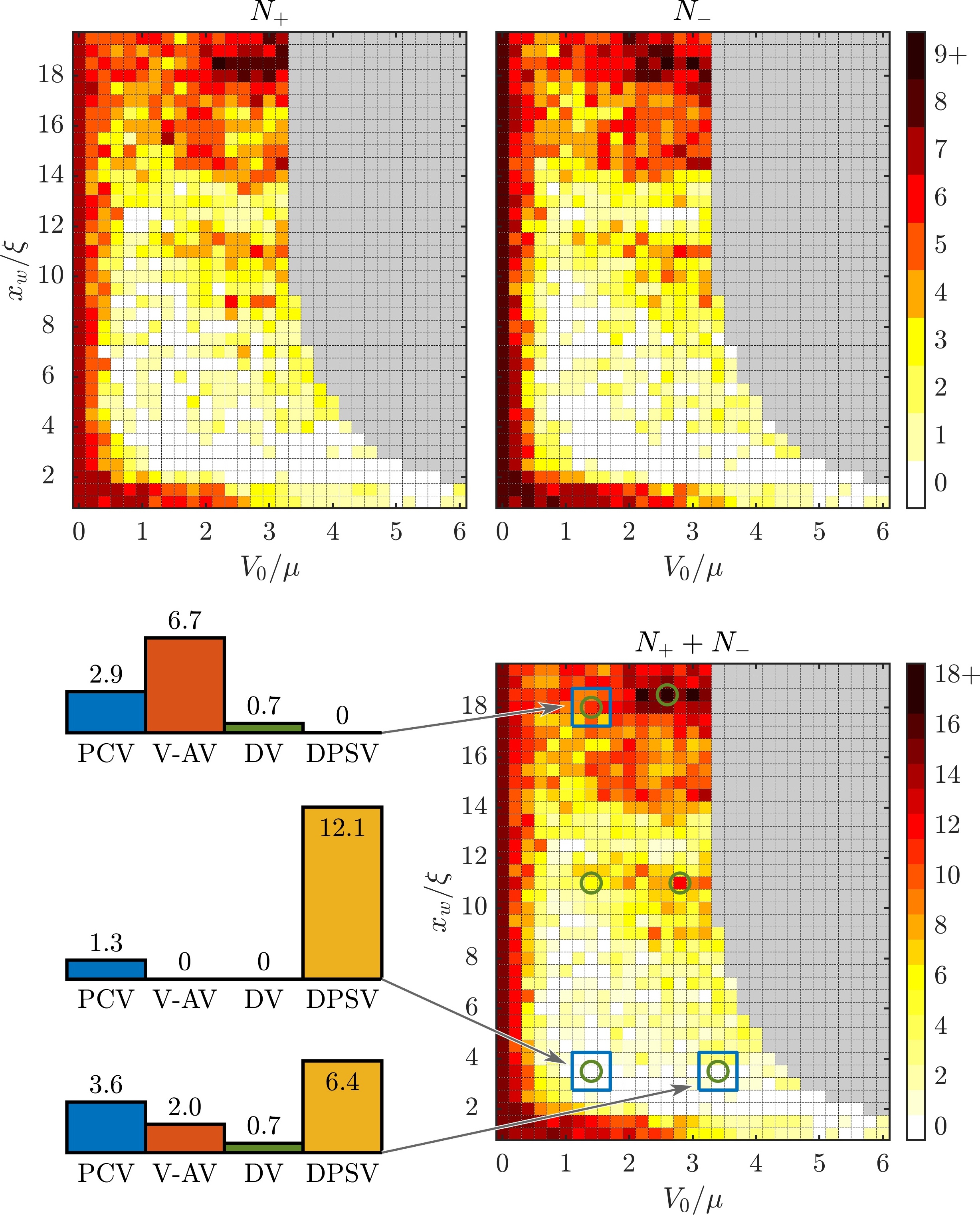}
    \caption{Loci of the number of remaining vortices after applying the OL combing process as a function of the OL parameters $V_0$ and $x_w$ corresponding, respectively, to its strength and fringe spacing. The system parameters correspond to $T_{\text{tot}} = 1000$~ms and $\gamma = 0.0005$, and $\beta = 2 \alpha = 2/5$ while other parameters (i.e., chemical potential and trapping strengths) are the same as in other figures. The top panels depict the total number of positively ($N_+$; left) and negatively ($N_-$; right) remaining vortices. The bottom-right panel displays the total number of remaining vortices ($N=N_++N_-$). The bottom-left panel depicts the breakdown of the average number of removed vortices averaged of the $3\times 3$ grid of $V_0$ and $x_w$ values corresponding to the indicated blue squares. The green circles indicate the locations for the six vortex count time series presented in Fig.~\ref{FigNnNp}.  The gray region corresponds to parameter values that we did not explore in detail as the OL strength is too strong and the condensate is fragmented into individual, basically, non-interacting condensates.
    }
    \label{FigLociT1000_NpNm}
\end{figure}

\begin{figure}
    \includegraphics[width=0.99\columnwidth]{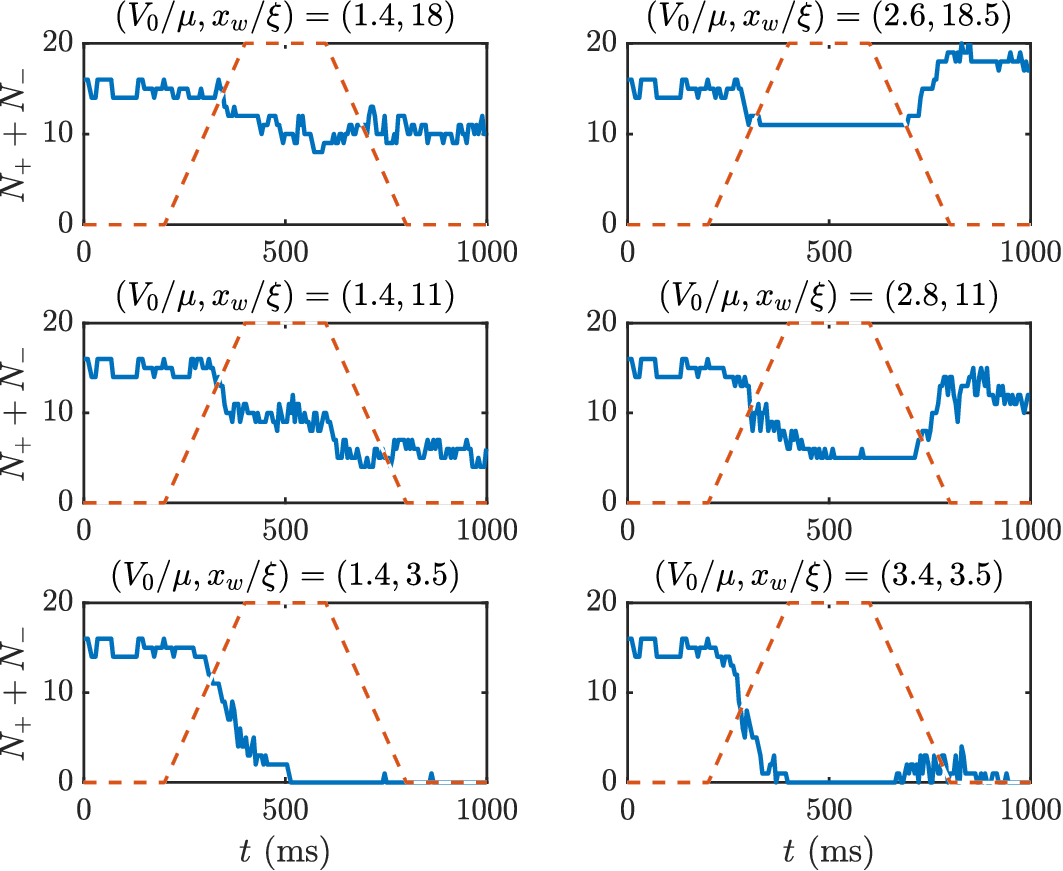}
    \caption{
    Time series of the combing removal for the six locations depicted by the green circles in Fig.~\ref{FigLociT1000_NpNm}. The red dashed line depicts the different stages of the comb ramping up and down.
    }
    \label{FigNnNp}
\end{figure}

Generally, we note the following outcomes of vortex elimination when varying $x_w$ and $V_0$:

\begin{enumerate}
\item {\em Role of $x_w$ and $V_0$ in vortex elimination by the PCV mechanism.} We observe that the PCV mechanism is ubiquitous in the entire region of interest as suggested in Fig.~\ref{FigLociT1000_NpNm}. PCV events seem to be most prevalent for (i) relatively thin fringes (of the order of 3--4 healing lengths) and relatively strong peak comb amplitudes (e.g., $3 \mu < V_0 < 4 \mu$) and for (ii) relatively wide fringes spacings ($x_w\approx 18 \xi)$ and weak peak comb amplitudes ($V_0\approx 1.5\mu$).
\item {\em Role of $x_w$ and $V_0$ in V-AV annihilations.}
V-AV annihilations are primarily observed for wide comb fringes with weaker amplitudes (e.g., $17.5 \xi < x_w < 18.5 \xi$ and $1.2 \mu < V_0 < 1.6 \mu$; see Fig.~\ref{FigLociT1000_NpNm}). It seems that fringes in this regime provide sufficient spacing for the vortices of opposite charge to become proximal and annihilate each other.

\item {\em Role of $x_w$ and $V_0$ in the DV vortex eliminations.}
We observe that vortex elimination by damping is only significant for cases of relatively high damping coefficients, and generally when other vortex elimination mechanisms are insignificant.  For experimentally relevant values of the damping coefficient, vortex removal due to damping is practically minimal compared with the other three vortex removal mechanisms.

\item {\em Role of $x_w$ and $V_0$ in the DPSV elimination mechanism.}
We note that the DPSV process is the dominant mechanism for vortex removal (over PCV, V-AV, and DV) in the regime of moderately thin fringes with sufficiently strong comb amplitudes (for instance, roughly in the region of the parameter space where $2.5 \xi <x_w<4.5 \xi$ and $1.4 \mu < V_0 <1.8 \mu$; see Fig.~\ref{FigLociT1000_NpNm}).  It appears that for a fixed $V_0$, smaller values of $x_w$ lead to larger potential energy and BEC density gradients, which introduce large variation of the nominal vortex characteristic length scale (the healing length) across the vortex. We suspect that such a large variation of background density over the size of the vortex may play a role in the causing the phase singularity of the vortex to get detached from the vortex core (defined by the density dip) and then ``lost'' into the vanishing density background at the periphery of the cloud.  Interestingly, during this process, the density excitation is reminiscent of a JR soliton confined to a narrow channel.  This soliton-like structure  appears to be generated (along with its associates phase profile) in the phase-singularity detachment process and subsequently move through the narrow channel; see the phase and vorticity profiles of the density dip of the DPSV in Fig.~\ref{Fig_DPSV} in comparison those of the vortex dipole and JR soliton in Fig.~\ref{Fig_JR}.

\end{enumerate}

\subsection{Optimized Protocols for Vortex Combing}
\label{sec:sub:optim}

Based on the above observations, we now assemble our conclusions regarding the protocols that yield the minimal number of remaining vortices. These findings are supported by two-parameter diagrams under variations of $x_w$ and $V_0$ in 
Figs.~\ref{FigLociT1000_NpNm}--\ref{FigLociAlphaBeta}, as we discuss in further detail in this section.  We also discuss the effect of variations of other parameters, including damping $\gamma$ and the parameters $\alpha$ and $\beta$ that regulate the comb ramp time scales, for a fixed total time $T_{\rm tot}$ of the ramping process.  For these studies, we map out in parameter space the overall efficiency of vortex elimination by summing $N_+$ and $N_-$, the numbers of of positively and negatively charged vortices, respectively, remaining at the end of the combing protocol including comb ramp-down.  
\begin{enumerate}
    
\item {\em Role of $x_w$ and $V_0$ in overall vortex removal.}
As depicted in Fig.~\ref{FigLociT1000_NpNm}, the removal of vortices depends non-trivially on the OL parameters. However, overall, there is a band of narrow widths of a few healing lengths where, for sufficiently large comb amplitudes (larger than one third or so of the condensate's chemical potential), most vortices are successfully removed.  Within that region, for relatively narrow fringe widths,  weaker comb amplitudes remove vortices more efficiently. Furthermore, for low comb amplitude, narrow fringes remove vortices more effectively than wider ones. Thus, taking into account that we would like the comb amplitude to be as weak as possible in order to minimize perturbations and excitations of the BEC, the optimal removal takes place at the lower-left corner of the locus (see region inside the bottom-left blue square in the $N_++N_-$ locus in Fig.~\ref{FigLociT1000_NpNm}). 

As we move from the lower right corner of the locus to the lower left, with a fixed fringe width, the PCV mechanism decreases in number slightly while the DPSV mechanism becomes more prominent. It is precisely the newly identified DPSV mechanism that is responsible for most of the vortex combing. However, it is important for $V_0$ to not be too low because then it has minimal influence on the vortex dynamics. Similarly, thinner fringes having a width on the order of the vortex core width strengthen the DPSV mechanism, while thicker fringes  enable the vortices to survive within a given fringe as if in a larger condensate.  Again, however, if the fringes become too thin, one obtains  an effective ``homogenized'' limit~\cite{sparber,ilan} where the vortices do not (practically) see the potential at all. In light of that, the fringes should not be too thin in order to efficiently remove vortices.

\begin{figure}
    \includegraphics[width=1\columnwidth]{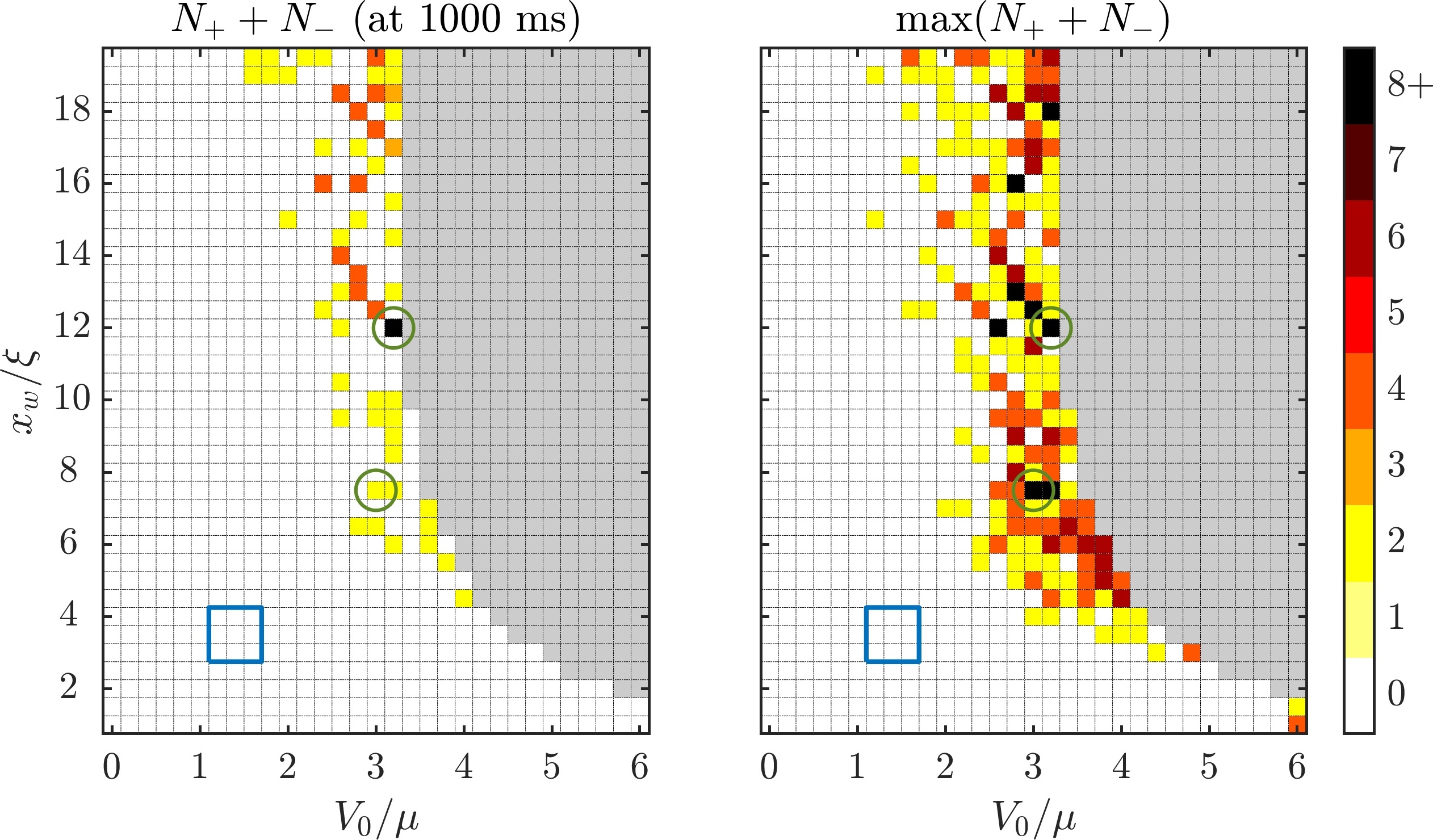}
    \\[2.0ex]
    \includegraphics[width=1\columnwidth]{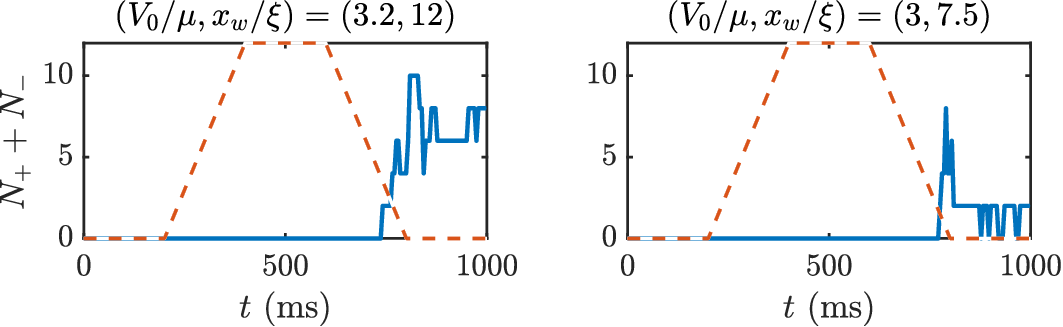}
    \caption{
    Vortices produced by the combing process when starting with no vortices before combing. The top left and right panels depict, respectively, the final and maximum number of vortices achieved during the combing process. The bottom two panels depict the time series of the vortex numbers for the two locations depicted by the green circles in the top panels.  The parameters are the same as in Fig.~\ref{FigLociT1000_NpNm}.
    }
    \label{FigLociNoStir}
\end{figure}

To supplement the previous results detailing the counts of the different vortex removal mechanisms, we depict in Fig.~\ref{FigNnNp} the time series of the number of vortices $N=N_++N_-$ as the combing process takes place for the six $(V_0,x_w)$-parameter locations indicated by the green circles in Fig.~\ref{FigLociT1000_NpNm}. As these different counts indicate, the efficient removal of vortices for relatively low values of $x_w$ degrades as $V_0$ increases past $V_0\approx 2\mu$ as spurious vortices are produced when ramping down the comb.  This effect is even more pronounced for $x_w$ values in the middle of the panel in Fig.~\ref{FigLociT1000_NpNm} where the comb effectiveness is dramatically reduced due to the production of vortices during the ramping down of the comb.  This unintended vortex production is due to the nucleation of vortices at the periphery of the cloud and is a consequence of the strong density gradient variations produced by relatively tall (large values of $V_0$) and thin comb fringes. This effect is more clearly elucidated in Fig.~\ref{FigLociNoStir} where the effects of the combing mechanism in the absence of initial vortices (namely, in the absence of the stirring process) is depicted. As the figure suggests, relatively large comb strengths result in steep and quickly changing density gradients that nucleate vortices. The above combined observations can be used to shed some light into the location of the observed optimum parameter region within the (wide) array of protocols explored in Figure~\ref{FigLociT1000_NpNm}.

\begin{figure}
    \includegraphics[width=1\columnwidth]{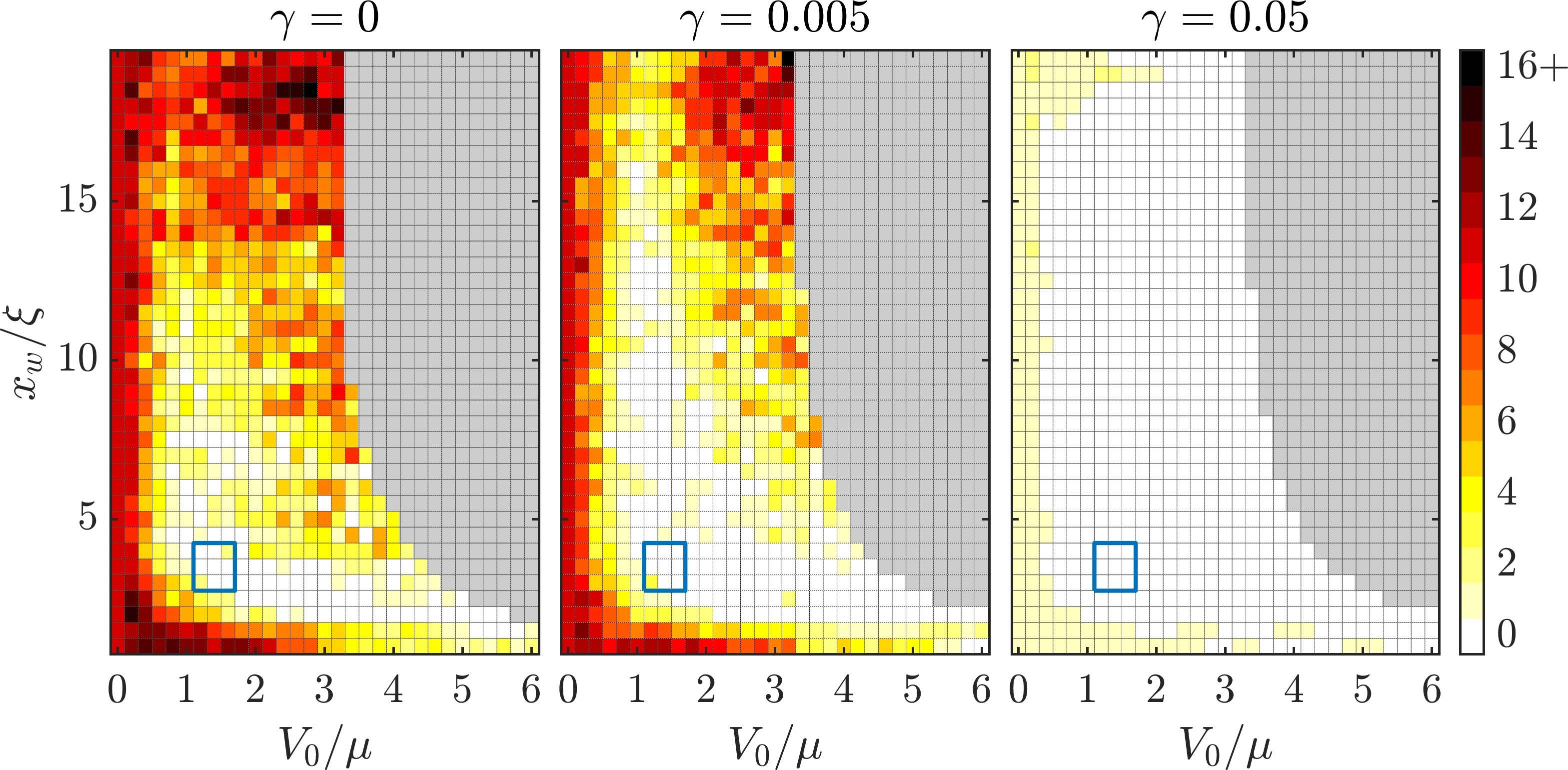}
    \caption{Loci of the total number of remaining vortices after combing for
    the following phenomenological dissipation values $\gamma = 0$, $\gamma = 0.005$, and $\gamma = 0.05$ (from left to right). All other parameters are the same as in Fig.~\ref{FigLociT1000_NpNm}.}
    \label{FigLociGamma}
\end{figure}

\item {\em Role of $\gamma$ in the remaining vortices.}
Stronger phenomenological dissipation ($\gamma$) leads to faster spiraling out of the vortices and thus a more effective DV removal. Figure~\ref{FigLociGamma} depicts the loci of remaining vortex after combing as $\gamma$ is increased (from left to right). As is observed in the figure, vortex removal by the dissipation mechanism (DV) does not contribute significantly for values of $\gamma$ of order of $10^{-3}$ or weaker. In this weak dissipation regime, which is precisely the regime of typical experiments, the comb removal mostly relies on the other removal mechanisms (PCV, V-AV, and DPSV). Therefore, for practical purposes and experimentally relevant situations, the combing is not significantly affected by the presence of thermal atoms for typical experimental temperatures of the BEC.

\begin{figure}
    \includegraphics[width=0.99\columnwidth]{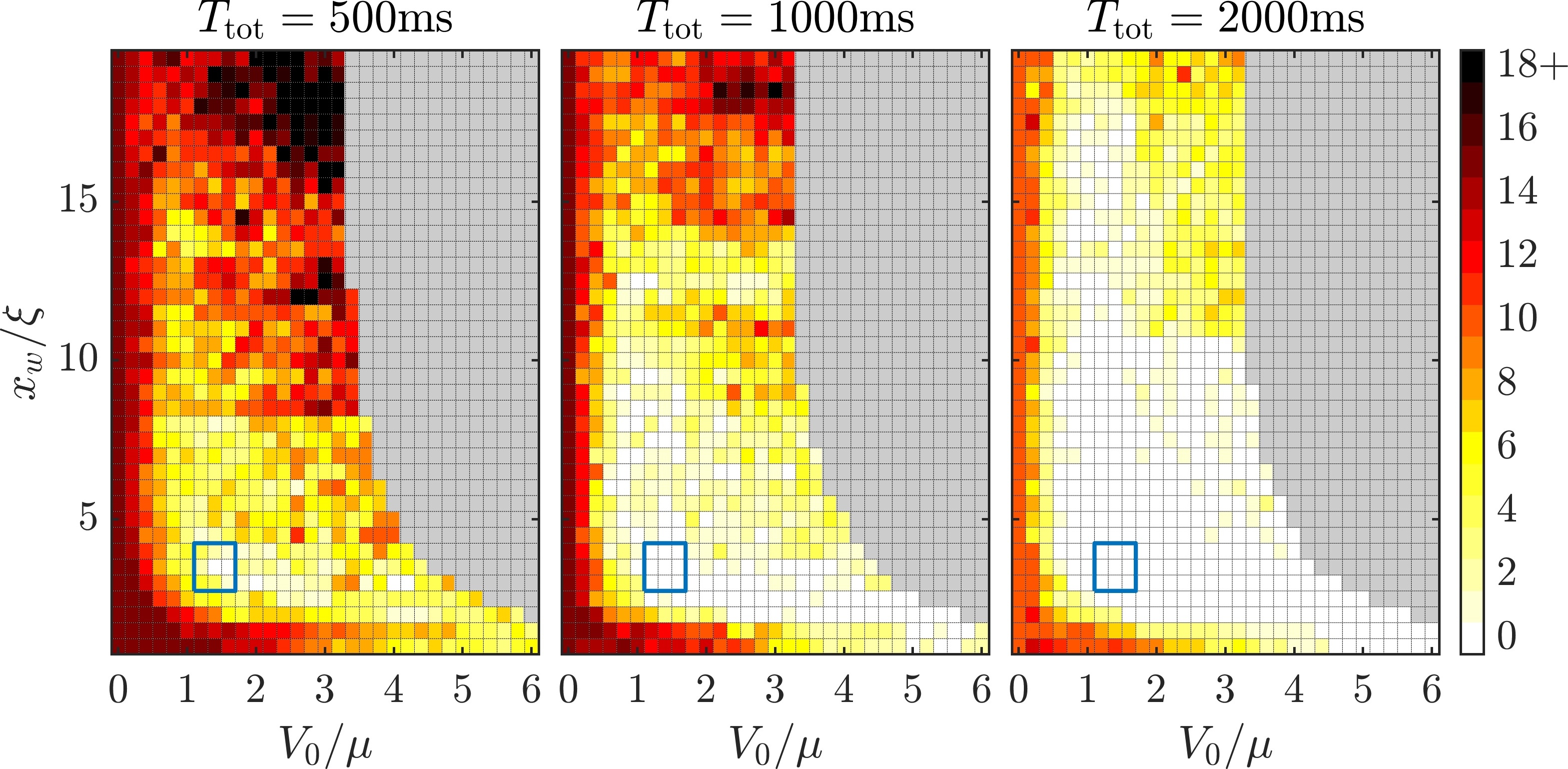}
    \caption{Loci of the total number of remaining vortices after combing for $T_{\text{tot}} = 500$, 1000, 2000~ms.  All other parameters are the same as in Fig.~\ref{FigLociT1000_NpNm}.}
    \label{FigLociTtot}
\end{figure}

\item {\em Role of $T_{\rm tot}$ in the remaining vortices.}
Naturally, the longer the OL is held on, the more the vortices are combed out of the condensate, as seen in  Figure \ref{FigLociTtot}.  All of the vortex removal mechanisms that we have identified, namely PCV, DPSV, AV, and DV, then get to act for a prolonged interval and affect the vortex dynamics. This helps comb the vortices out of the BEC cloud when the OL is switched on for a longer time.

\begin{figure}
    \includegraphics[width=0.6\columnwidth]{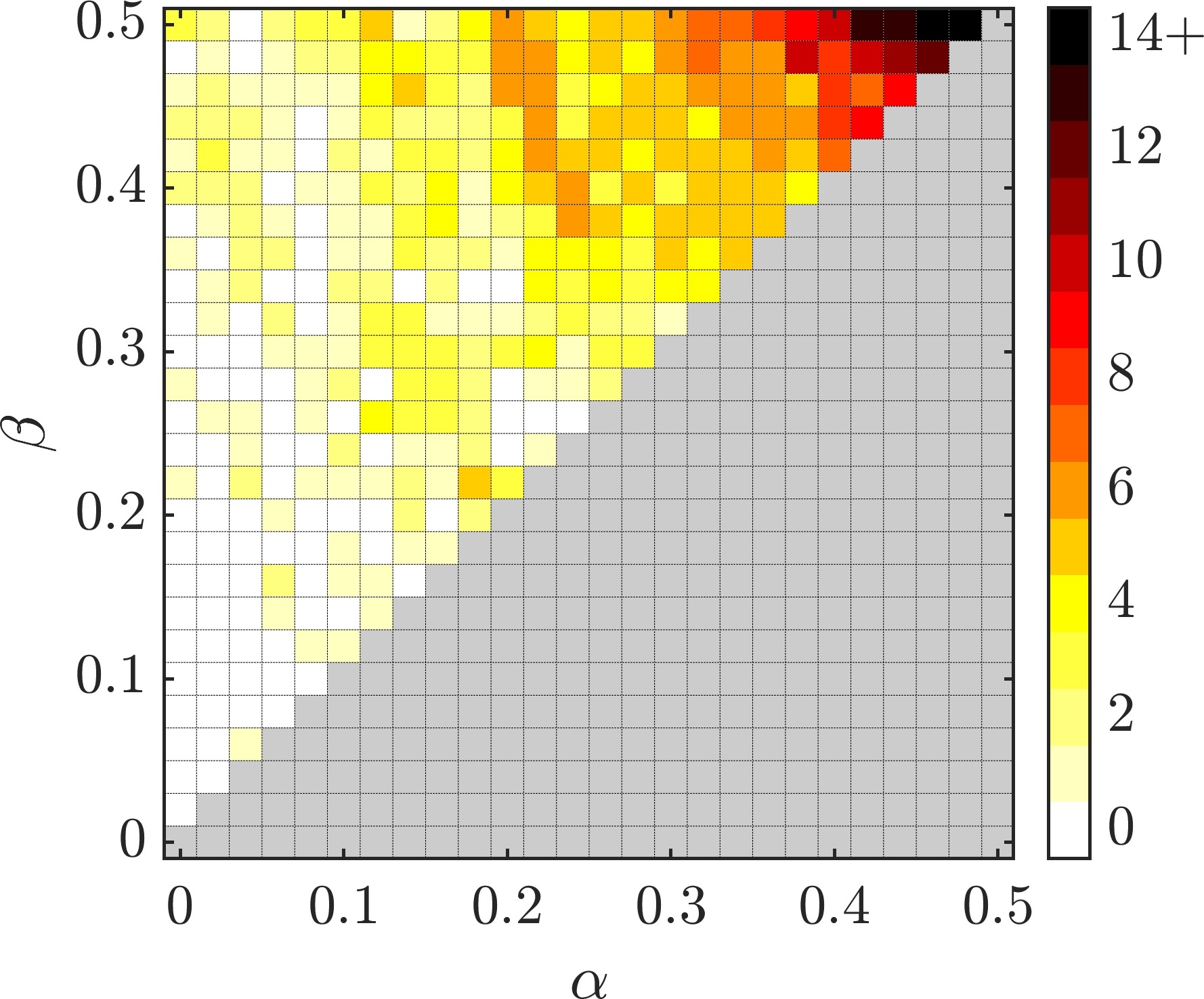}
    \caption{Loci of the total number of remaining vortices for different values of $\alpha$ and $\beta$ that control the ramping up, holding, and ramping down of the combing process (see Fig.~\ref{FigHoldProfile}).  All other parameters are the same as in Fig.~\ref{FigLociT1000_NpNm}. The region in gray is forbidden as, by definition, $\beta>\alpha$.}
    \label{FigLociAlphaBeta}
\end{figure}

\item {\em Role of the ramping up, holding, and ramping down intervals.}
Figure~\ref{FigLociAlphaBeta} shows the remaining vortex counts as the holding profile parameters $\alpha$ and $\beta$ are varied (see Fig.~\ref{FigHoldProfile}). Not surprisingly, for smaller values of $\alpha$ and $\beta$, corresponding (for a constant total combing time $T_{\rm tot}$) to a longer holding time of the comb, the combing mechanism is more efficient at removing vortices.  Interestingly, there seems to be an overall tendency for the number of removed vortices to be approximately constant for  $\alpha+\beta= {\rm const}$. In hindsight, this is straightforward since the total combing ``work'' or ``action'' could be quantified  by using the integral under the combing profile which is proportional to $A=1-(\alpha+\beta)$ (area under the curve of Fig.~\ref{FigHoldProfile}). Thus, constant $\alpha+\beta$ is tantamount to a family of holding profiles with approximately the same combing action (when combining ramping up, holding, and ramping down intervals).

\item {\em The optimality of the vortex removal is robust with respect to variations on the number of atoms and the geometry of BEC cloud.} 
Finally, to complement our results, Fig.~\ref{FigLoci2muCircle} shows the combing efficiency for different atom numbers and geometries of the confining BEC trapping. Specifically, the left panel of Fig.~\ref{FigLoci2muCircle} shows the remaining vortex count locus for a BEC cloud prepared with a larger number of atoms corresponding to twice the chemical potential compared to our previous results ($\mu=2\!\times\! 5.2844 \, \hbar \omega_z$). As the figure suggests, the location of the optimal combing parameters for this larger chemical potential case is, as before, around the lower-left corner of the parameter space. This seems to suggest that (in the region of parameters of interest) optimal combing is not highly sensitive with respect to the number of atoms in the BEC or the chemical potential.

\begin{figure}
    \includegraphics[height=3.7cm]{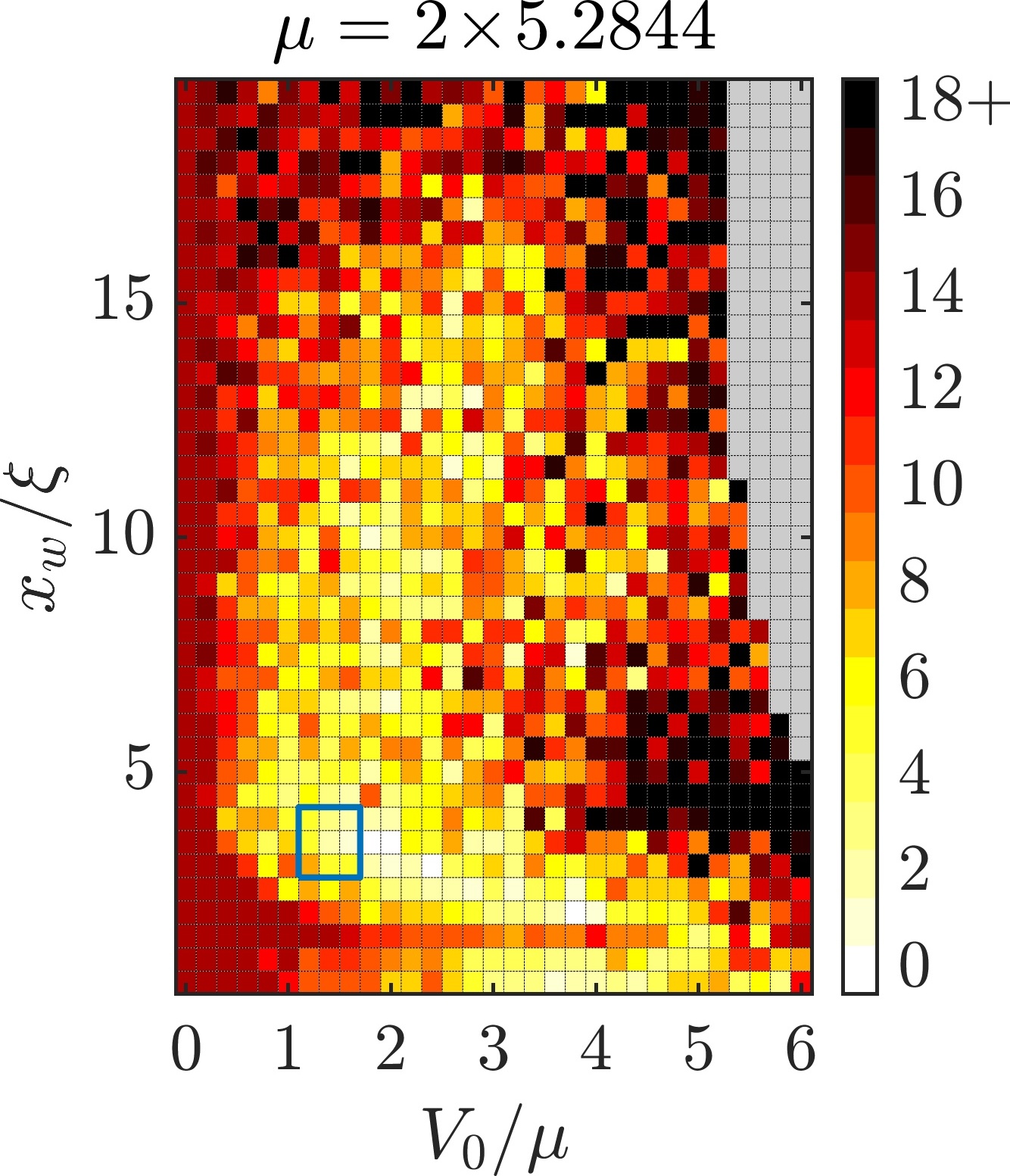}
    \includegraphics[height=3.7cm]{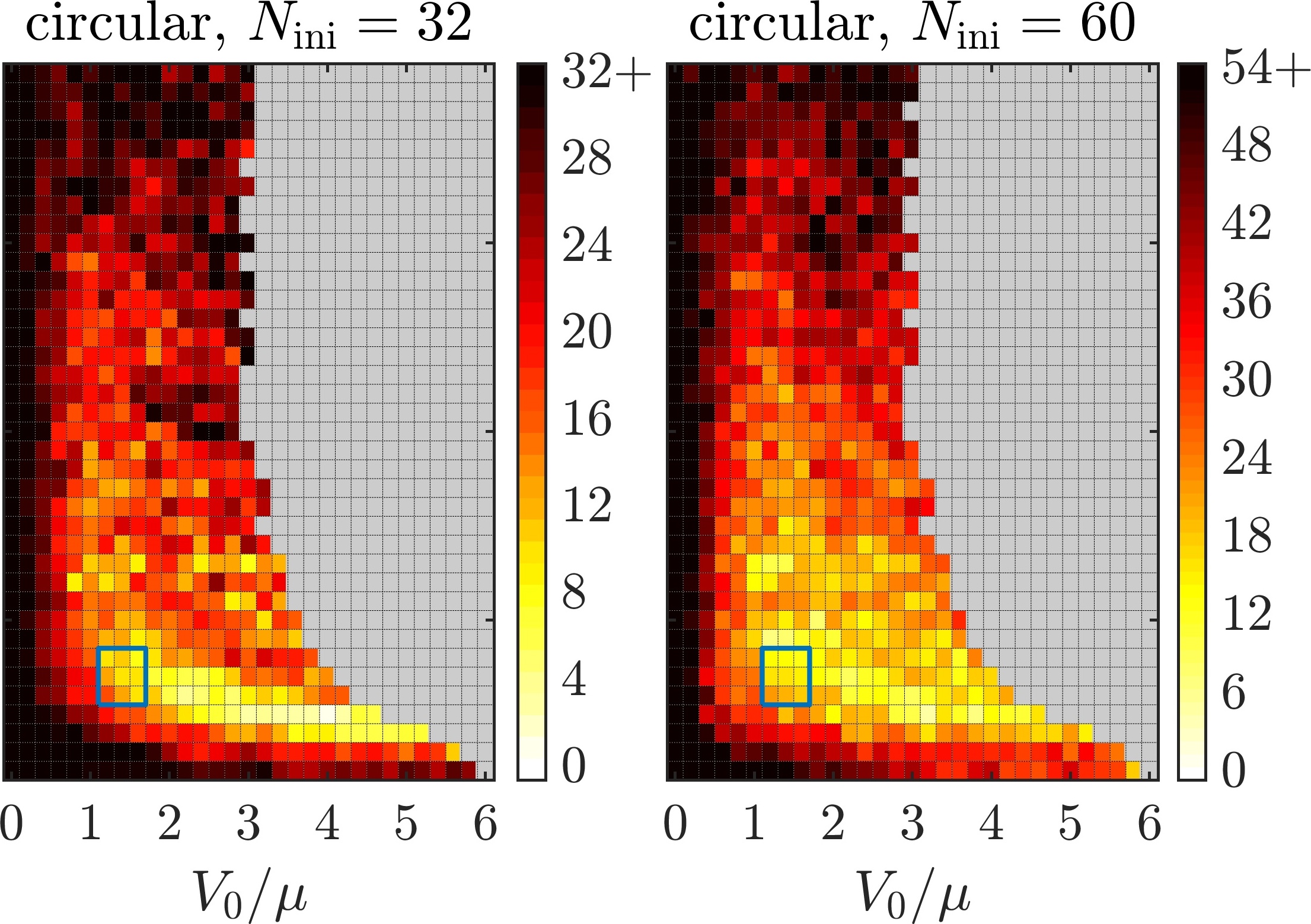}
    \caption{Left panel: same as the left panel of Fig.~\ref{FigLociTtot} but for twice the value of $\mu$ as previously used, so that now $\mu=2\!\times\! 5.28 \, \hbar \omega_z$. Middle and right panels: same as Fig.~\ref{FigLociTtot} but for a circular trap with trapping strengths $(\omega_x,\omega_y,\omega_z) = 2\pi (8.5,8.5,38.5)$Hz. The middle and  right panels correspond, respectively, to starting with a total vortex number of 32 and 60. The latter case corresponds to an initial vortex density similar to the one in the elliptical trap cases.}
    \label{FigLoci2muCircle}
\end{figure}

We also explored the effects of changing the geometry of the cloud. Specifically, the middle and right panels of Fig.~\ref{FigLoci2muCircle} depict the remaining vortex count loci for a {\em circular} trap. The intention of this set of results is twofold.  First, to compare the previous case of an elliptical trap ---where the vortices had a narrower vertical distance to travel to the edge of the cloud to get combed out--- with the circular trap case where vortices need to travel a longer distance for them to be eliminated at the edge of the condensate (via the DPSV or PCV mechanisms). Second, the middle and right panels of Fig.~\ref{FigLoci2muCircle} correspond to different initial vortex densities (the right panel has about twice as many vortices compared to the middle panel) before the combing process is turned on. The results tend to suggest that the overall shape of the loci is not highly dependent on the number of vortices to be combed. Nonetheless, it is important to mention that, for larger chemical potentials and for larger number of vortices, complete vortex removal would need longer combing times ---note that the remaining vortex count minima of the loci  in Fig.~\ref{FigLoci2muCircle} are not zero.
In that sense, the elliptical, anisotropic trap
setting is more prone to combing than its isotropic,
circular analogue.
Overall, the results of Fig.~\ref{FigLoci2muCircle} tend to suggest that, as was the case before, the optimal combing parameters lie in the bottom-left corner of the parameter space.\\

\end{enumerate}

\section{Conclusions and Future Exploration}
\label{sec:conclu}
In this work, we have experimentally demonstrated a novel method for efficiently removing vortices from a highly oblate BEC, and we have studied the vortex removal mechanisms in depth with simulations based on the 2D GPE.   Our vortex removal technique relies on the application of an optical lattice potential to the BEC for a short period of time. Our main findings indicate that for the parameter ranges studied, there exists an optimal range for optical lattice depth and periodicity for efficient vortex elimination from the condensate.  Within this optimal range, vortex removal occurs via a variety of mechanisms, including by a mechanism that involves the density modulation location of the typical vortex profile separating from the location of the phase singularity.  We are unaware of previous studies or descriptions of this dynamical process.  Generally, the ability to efficiently remove vortices from a BEC may prove useful in experiments for which the BEC must be in its lowest energy state prior so subsequent experimental protocols, such as in controlled studies of vortex nucleation and dynamics at the few-vortex level; relevant studies that could make use of this technique have been experimentally performed, e.g., in Refs.~\cite{dsh1,bagnato,dsh2}.  The mechanisms investigated here might also be used for studies of vortex dynamics in turbulent systems, perhaps for which vortices are removed from a section of the BEC by a combing technique, and subsequent dynamics of the remaining vortices are examined.  

The considerations presented herein suggest numerous additional possibilities for future work.  The systematic characterization of the DPSV mechanism and its specific conditions for existence is of particular interest. The connection of this mechanism, among others, with the existence and stability of JR soliton (and also vortex-dipole) traveling states {\it in narrow channels} is also of considerable interest. Finally, we remain mindful that all of our computational considerations were constrained (due to the computational expense of 3D runs) to two spatial dimensions. Nevertheless, it is of particular interest to characterize more systematically the dynamics of the associated narrowly confined vortex lines in three spatial dimensions; see, e.g., the visualizations of such confined vortical structures (in a rotating condensate) in the work of Ref.~\cite{ionutd}. These topics may be explored in future work.


\bibliographystyle{apsrev4-1}
\bibliography{vortex_comb_all_refs}

\end{document}